# Dynamics of Two Higgs Doublet CP Violation and Baryogenesis at the Electroweak Phase Transition

James M. Cline

*McGill University, Montréal, PQ H3A 2T8, Canada,*

Kimmo Kainulainen

*CERN, CH-1211, Geneve 23, Switzerland,*

and

Axel P. Vischer[1]

*School of Physics and Astronomy, University of Minnesota*

*Minneapolis, MN 55455, USA.*

## Abstract

We quantitatively study the charge transport mechanism of electroweak baryogenesis in a realistic two-Higgs-doublet model, comparing the contributions from quarks and leptons reflecting from electroweak domain walls, and comparing the exact profile of the CP-violating phase with a commonly used ansatz. We note that the phenomenon of spontaneous CP violation at high temperature can occur in this model, even when there is no CP violation at zero temperature. We include all known effects which are likely to influence the baryon production rate, including strong sphalerons, the nontrivial dispersion relations of the quasi-particles in the plasma, and Debye screening of gauged charges. We confirm the claim of Joyce, Prokopec and Turok that the reflection of tau leptons from the wall gives the dominant effect. We conclude that this mechanism is at best marginally capable of producing the observed baryon asymmetry of the universe, and we discuss some ways in which it might be enhanced.

---

[1]Current address: Niels Bohr Institute, DK-2100, Copenhagen Ø, Denmark.

# 1 Introduction

An exciting proposal during the last few years is that the baryon asymmetry of the universe was created during the electroweak phase transition [1], by harnessing both the anomalous baryon number violation present within the standard model, and the first order nature of the phase transition which gives the necessary departure from thermal equilibrium. The third ingredient needed for baryogenesis is CP violation, which is widely believed to be too small for this purpose in the standard model. However in modest extensions of the standard model it is possible to introduce new sources of CP violation which are more effective for baryogenesis [2] - [5]. For example in generic models with more than one Higgs doublet, phases can be introduced into the potential for the scalars which are only weakly constrained by laboratory limits on CP violation [6].

The simplest example is a theory with two Higgs doublets, where there is a single phase $\theta$, namely the phase mismatch between the VEV's of the two scalar fields [2]. Although at zero temperature $\theta$ is just another parameter of the theory, during the electroweak phase transition it is a spatially varying field $\theta(x)$ whose value depends on the position relative to the domain walls that separate the true and false vacuum phases during the transition. In principle, the detailed form of $\theta(x)$ is needed for computing the difference in reflection probabilities between quarks and antiquarks bouncing off the domain walls, which is how a baryon asymmetry is produced in the charge transport mechanism of electroweak baryogenesis [7, 8]. Until now there have been no attempts to compute $\theta(x)$; rather an ansatz has been invoked. We will show that the difference between the actual solutions and the ansatz can be quite significant. Moreover, a detailed study of the equation of motion for $\theta(x)$ reveals the interesting phenomenon of spontaneous CP violation [9] at high temperature that can occur in this model, even when there is no CP violation at zero temperature.

We subsequently use our solution for $\theta(x)$ to study the baryon production in the charge transport mechanism. We account for all known effects which are likely to influence the baryon production rate including the nontrivial dispersion relations of the quasiparticles in the plasma, strong sphalerons and Debye screening of gauged charges. We do not include thermal damping in our calculation of reflection amplitudes, but we will argue that this phenomenon is likely to be of little consequence to our results. We confirm the claim of Joyce, Prokopec and Turok [8] that the reflection of tau leptons from the wall gives the dominant effect; however we differ from ref. [8] in that our exact computation of the probabilities for particles to reflect from the bubble wall gives smaller results than their estimates. Based on our results, it appears difficult to get a large enough baryon asymmetry even using the most favorable values of large CP violation and slow bubble wall velocity. However some modifications, such as a larger tau lepton Yukawa coupling or further slowing of the bubble walls toward the end of the phase transition, may make it possible to account for the baryon asymmetry with this mechanism.

In section 2 we introduce a two-Higgs-doublet model which is sufficiently realistic to include the physics needed for electroweak baryogenesis, and we find its corresponding finite-temperature potential. Section 3 discusses the solution for the $\theta$ field equation of motion, and in section 4 the computation of the fermion reflection coefficients is explained. Section 5 treats the complicated process of how the fermions reflected from the wall will propagate back into the symmetric phase before the wall overtakes them again. In section 6 we put these results together to find the baryon asymmetry. We present our results and conclusions in section 7.

## 2 The Model and its Effective Potential

In the most general potential with two Higgs doublets, flavor-changing neutral currents (FCNC's) are unsuppressed; it is therefore convenient to impose a global symmetry such as $\Phi_1 \to -\Phi_1$ to forbid them [10]. This symmetry, if exact, would also forbid CP-violation and lead to domain wall formation in the early universe, but these problems can be cured without reintroducing the first by allowing the symmetry to be softly broken by a term $\Phi_1^\dagger \Phi_2$ [2, 11]. The potential is

$$\begin{aligned}
V(\Phi_1, \Phi_2) &= -\mu_1^2 \Phi_1^\dagger \Phi_1 - \mu_2^2 \Phi_2^\dagger \Phi_2 + \tilde{\kappa} \Phi_1^\dagger \Phi_2 + \tilde{\kappa}^* \Phi_2^\dagger \Phi_1 + \frac{\lambda_1}{2}(\Phi_1^\dagger \Phi_1)^2 + \frac{\lambda_2}{2}(\Phi_2^\dagger \Phi_2)^2 \\
&\quad + h_1(\Phi_1^\dagger \Phi_2)(\Phi_2^\dagger \Phi_1) + h_2(\Phi_1^\dagger \Phi_1)(\Phi_2^\dagger \Phi_2) + h_3\left((\Phi_1^\dagger \Phi_2)^2 + (\Phi_2^\dagger \Phi_1)^2\right) \\
&\quad + y_2 \overline{F}_L \Phi_2 f_R + \text{ c.c.}
\end{aligned} \qquad (1)$$

The last term is the Yukawa interaction of the Higgs field with a generic fermion field, taking into account that the coupling $y_1$ to $\Phi_1$ is forbidden by the global symmetry.

In the above expression we have used global field redefinitions to make the coupling $h_3$ real. In general the mass term $\tilde{\kappa}$ thus remains complex. If $h_3$ should happen to be zero, as is the case at tree level in the supersymmetric (SUSY) standard model, the same redefinition could be used to make $\tilde{\kappa}$ real and then there would be no CP-violation in the Higgs sector. However even in the SUSY case a complex value of $h_3$ is generated at one loop [7]. We will write $\tilde{\kappa}$ as a modulus times a phase,

$$\tilde{\kappa} = \kappa e^{i\delta_\kappa}. \tag{2}$$

Finite-temperature corrections to the effective potential at the one-loop level are given by the integral [12]

$$\Delta V_T(\phi) = \mp \frac{T^4}{2\pi^2} \int_0^\infty dx\, x^2 \operatorname{Tr} \ln\left(1 \pm \exp[-\sqrt{x^2 + (M(\phi)/T)^2}\,]\right), \tag{3}$$

where the trace is over all particles in the theory and $+(-)$ is for fermions (bosons). Each real field (four for a Dirac fermion) counts as a single state in the sum. The masses are evaluated at arbitrary background values of the scalar fields. Expanding $\Delta V_T(\phi)$ to fourth order in the masses, the result can be written as

$$\Delta V_T(\phi) \cong \sum_i \left\{ C_i \frac{m_i^2 T^2}{48} - D_i \frac{m_i^3 T}{12\pi} - E_i \frac{m_i^4}{64\pi^2}(\ln(m_i^2/T^2) - F_i) \right\} \tag{4}$$

where $C_i = 2(1)$, $D_i = 1(0)$, $E_i = 1(-1)$ and $F_i \cong 5.41(2.64)$ for bosons (fermions). For example, the quadratic term has the effect of shifting the parameters $\mu_i^2$ of the tree-level Lagrangian to

$$\mu_i^2(T) = \mu_i^2 - a_i T^2 \tag{5}$$

where in terms of the SU(2) and U(1) gauge couplings and the Yukawa couplings,

$$a_i = \frac{1}{12}(3\lambda_i + h_1 + 2h_2) + \frac{1}{16}(3g^2 + g'^2) + \frac{1}{4}y_i^2. \tag{6}$$

(Note that $y_1 = 0$ for our model). The ring-improvement of the potential is the first iteration of eq. (3), in which the masses $M$ of the bosons are taken to be those at tree-level plus the one-loop temperature-corrected ones [13].

The effective potential derived here differs from that of Turok and Zadrozny [2] who considered the same model without the $\kappa$ term. Their computation was made in the unitary gauge, which gives unreliable results, as has been shown in ref. [14]. One way to see that the result of [2] is incorrect is the fact that their thermal corrections do not respect the symmetry $\Phi_1 \leftrightarrow \Phi_2$ even though the underlying Lagrangian does. Ref. [14] shows that unitary gauge is a poor choice near the critical temperature, where each order in the loop expansion is as important as the next for obtaining quantities that are perturbatively calculable in covariant gauges.

The phase transition in this model generically proceeds in two stages, because there are separate critical temperatures for the two fields [15, 16]. To sidestep this complication we now follow previous authors by making the simplifying technical assumption that $\lambda_1 = \lambda_2 \equiv \lambda$, $\mu_1^2 = \mu_2^2 \equiv \mu^2$ [2, 17]. Then the Higgs potential is invariant under the exchange of the moduli of the two fields, and we can parametrize the domain walls separating the true and false vacua in the form

$$\Phi_1(x) = \frac{1}{\sqrt{2}} \begin{pmatrix} 0 \\ \rho(x) e^{i(\alpha(x)+\theta(x))/2} \end{pmatrix}, \qquad \Phi_2(x) = \frac{1}{\sqrt{2}} \begin{pmatrix} 0 \\ \rho(x) e^{i(\alpha(x)-\theta(x))/2} \end{pmatrix}. \qquad (7)$$

The CP-conserving phase $\alpha(x)$ is the neutral Goldstone boson which is eaten by the $Z_0$ when the symmetry is broken. Its expectation value is a constant which can be ignored in our subsequent discussion; therefore the phase transition is described by two real fields instead of three. Moreover in a first approximation, the phase $\theta(x)$ can be treated as a small perturbation, so that the domain wall profile during the phase transition is determined by a single equation for $\rho(x)$. We expect the physics of quark reflection from bubble walls in this model to be similar to that of the more realistic case when $\mu_1^2 \neq \mu_2^2$.

It is clear from the full potential that if, for small $\theta(x)$, we want the symmetry to break in the direction of $\Phi_1 = \Phi_2$, then we must demand that

$$\kappa < 0. \qquad (8)$$

If not, the same physics would still ensue except that we would have to change the name of one of the fields, say $\Phi_1 \to -\Phi_1$; thus we will take (8) as our convention for the sign of $\kappa$.

Unfortunately the necessity of coupling the fermions to only one Higgs field means that finite-temperature corrections will spoil the symmetry that would allow both fields to have equal VEV's as in (7). We will ignore this complication in order to maintain the single stage phase transition, yet still keep the effect of fermion contributions to the finite-temperature effective potential, using the prescription that

$$\mu^2(T) = \mu^2 - aT^2; \qquad a = (a_1 + a_2)/2. \tag{9}$$

This should be regarded as a reasonable compromise between realism and simplicity. It would be exact in the case where the fermions coupled with equal strength to both Higgs doublets, which is another, less familiar way of avoiding flavor-changing neutral currents (for further discussion of this point, see section 7). We will keep only the dominant top quark Yukawa contribution $y_2 = 1.4$ in $a_2$. This value corresponds to a mass of 176 GeV [18].

The result of substituting the form (7) into the effective potential for the Higgs fields is

$$\begin{aligned} V_{\text{eff}}(\rho, \theta) &= (-\mu^2 + aT^2 + \kappa \cos(\theta - \delta_\kappa))\rho^2 \\ &\quad - \delta T \rho^3 + \frac{1}{4}(\lambda_{\text{eff}} + 2h_3(\cos(2\theta) - 1))\rho^4 + \cdots \\ \lambda_{\text{eff}} &= \lambda + h_1 + h_2 + 2h_3. \end{aligned} \tag{10}$$

The ellipsis represents contributions which have a smaller effect on the evolution of the fields than those retained: temperature corrections to the quartic couplings, and cubic terms of the form $(\rho^2 + cT^2)^{3/2}$. The cubic term retained in (10) is contributed by the transverse gauge bosons whose thermal squared mass (the $cT^2$ term) vanishes at order $g^2$. They give $\delta = (2g^3 + (g^2 + g'^2)^{3/2})/(12\sqrt{2}\pi) = 0.018$. However this is an underestimate since the $(\rho^2 + cT^2)^{3/2}$-like terms must still function in somewhat the same way as a pure cubic term, and they are numerous because the two Higgs doublets contribute a total of eight particles in the sum (4). These terms are inconvenient to include explicitly because they make it impossible to compute the parameters of the phase transition analytically. Instead we will parametrize their effect, as well as contributions from other possible particles such as singlet Higgs fields [19] and possible nonperturbative effects [20], by keeping $\delta$ as an adjustable parameter (for more discussion, see section 7). In any case, this is necessary for avoiding the problem of baryon washout in the broken phase [17]. To see this, we note that the phase transition occurs at the critical temperature

$$T_c^2 = (\mu^2 - \kappa)/(a - \delta^2/\lambda_{\text{eff}}) \tag{11}$$

defined as the temperature when two degenerate minima develop in the Higgs potential. At this time the VEV is

$$\rho_c = 2\delta T_c/\lambda_{\text{eff}}. \tag{12}$$

Residual sphaleron interactions in the broken phase can wash out the baryon asymmetry once it has been created unless the condition $\rho_c \gtrsim T_c$ is satisfied, or in other words $\delta \gtrsim \lambda_{\text{eff}}/2$. (This can be derived from the condition that the sphaleron energy $E_{\text{sph}}$ be greater than $\approx 45 T_c$ [3] using $E_{\text{sph}} = 8\pi B(\lambda_{\text{eff}}) M_W(T)/g^2$, where $B(\lambda_{\text{eff}}) \simeq 1.6$ for small values of $\lambda_{\text{eff}}$.) Unfortunately the value $\delta = 0.018$ due to transverse gauge bosons would necessitate too small a value of $\lambda_{\text{eff}}$ to be compatible with the laboratory bound of $m_{h^0}^2 = \lambda_{\text{eff}} \rho_0^2 > (60 \text{ GeV})^2$. Henceforth we will therefore take $\delta = \lambda_{\text{eff}}/2$.

For future reference we give the VEV of the $\rho$ field and the vacuum masses of the Higgs fields here, in the limit of a small CP-violating phase. At zero temperature the potential is minimized by

$$\rho_0 = (2(\mu^2 - \kappa)/\lambda_{\text{eff}})^{1/2} = \frac{246}{\sqrt{2}} \text{ GeV} \tag{13}$$

and if we define a parameter

$$\zeta = -\kappa/m_{h^0}^2; \qquad \kappa = \frac{\mu^2}{1 - (2\zeta)^{-1}}, \tag{14}$$

the masses can be written as

$$\begin{aligned}
m_{h^0}^2 &= \lambda_{\text{eff}} \rho_0^2 \\
m_{A^0}^2 &= -4h_3 \rho_0^2 - 2|\kappa| = m_{h^0}^2 (2\zeta - 4h_3/\lambda_{\text{eff}}) \\
m_{H^0}^2 &= (\lambda_{\text{eff}} - 2h_1 - 2h_2 - 4h_3) \rho_0^2 - 2|\kappa| \\
&= m_{h^0}^2 (1 + 2\zeta - (2h_1 + 2h_2 + 4h_3)/\lambda_{\text{eff}}) \\
m_{H^\pm}^2 &= -(h_1 + 2h_3) \rho_0^2 - 2|\kappa| \\
&= m_{h^0}^2 (2\zeta - (h_1 + 2h_3)/\lambda_{\text{eff}}).
\end{aligned} \tag{15}$$

Note that $h^0$ is the Higgs field which gets a VEV. To explore the implications of various choices of the parameter $\kappa$, it will be helpful to invert the relations (15) to solve for the quartic couplings as functions of $\zeta$ and the mass ratios

$$\gamma_i \equiv m_i^2/m_{h^0}^2, \tag{16}$$

with $\gamma_{h^0} \equiv 1$. The couplings can then be written as

$$\begin{aligned}
\hat{h}_1 &\equiv h_1/\lambda_{\text{eff}} = (2\zeta - 2\gamma_{\pm} + \gamma_A)/2; \\
\hat{h}_2 &\equiv h_2/\lambda_{\text{eff}} = (1 - 2\zeta + 2\gamma_{\pm} - \gamma_H)/2; \\
\hat{h}_3 &\equiv h_3/\lambda_{\text{eff}} = (2\zeta - \gamma_A)/4.
\end{aligned} \quad (17)$$

From these we can express the parameter $a$ of eq. (9) in terms of the mass parameters,

$$a = \lambda_{\text{eff}} \left( \frac{1}{6} - \frac{\zeta}{3} + \frac{1}{24} \sum_{\substack{h^0,A,H \\ +,-}} \gamma_i \right) + \frac{3g^2 + g'^2}{16} + \frac{y^2}{8} \quad (18)$$

which is useful for determining the critical temperature through eq. (11), and thereby the width of the bubble wall to be discussed below.

Curiously, the combination $a - \delta^2/\lambda_{\text{eff}}$ can vanish for sufficiently large negative values of $\kappa$ (large positive $\zeta$), the mass which mixes the two Higgs fields in the tree-level potential. From eq. (11) we see that this would imply a negative value for $T_c^2$, meaning that there would be no phase transition. We should not trust our effective potential for these parameters, (and even if we could, there would be no electroweak baryogenesis), so we will exclude this region. Using the top quark Yukawa coupling $y = 1.4$, one can solve for the condition that there is a phase transition,

$$\begin{aligned}
\zeta &< 3c/\lambda_{\text{eff}} + 1/2 - 3(\delta/\lambda_{\text{eff}})^2 + \sum_{\substack{h^0,A,H \\ +,-}} \gamma_i/8 \\
c &= \frac{3g^2 + g'^2}{16} + \frac{3y^2}{8} = 0.34
\end{aligned} \quad (19)$$

For the smallest experimentally allowed values of $m_{h^0}^2$, $\lambda_{\text{eff}} = 0.12$ and assuming that $\delta = \lambda_{\text{eff}}/2$ and the mass ratios $\gamma_i = 1$ we get $\zeta < 8.8$. This condition can be violated only if there is a fine-tuned cancellation between $\mu^2$ and $\kappa$ designed to keep the weak scale at 100 GeV. For heavy Higgs boson masses the restriction on $\zeta$ becomes even less severe.

## 3  Bubble Wall Profiles

We now turn to the description of the domain wall at the phase transition. In principle the equations of motion for the bubble wall nonlinearily couple all doublet field components.

For technical reasons we will assume that the relative phase between the Higgs doublet is small, and therefore we work only to the lowest order in $\theta(x)$; in this case the equations for the modulus $\rho$ decouples from that of $\theta$. The equation for $\theta$ has nontrivial solutions because of the explicit CP-violation present in the effective action in the form of a complex $\kappa$. Moreover, we find that for a certain region of parameters CP gets spontaneously broken in the symmetric phase, giving raise to nontrivial solutions even when the effective action has no explicit CP-violation.

Since the bubbles of true vacuum grow to macroscopic proportions one can approximate the walls as planar, depending only on a single coordinate $z$ [21]. Combining the effective potential (10) with the kinetic energy density

$$E_{\text{kin}} = (\partial_z \rho)^2 + \frac{1}{4}(\rho \partial_z \theta)^2, \tag{20}$$

one finds that to zeroth order in $\theta$, the equation of motion of the $\rho$ field has the usual kink solution

$$\begin{aligned} \rho(z) &= \rho_c g(z); \\ g(z) &= \frac{1}{2}\Big(1 + \tanh(z/\Delta_{\text{wall}})\Big); \\ \Delta_{\text{wall}} &= \frac{4\rho_0}{m_{h^0}\rho_c}. \end{aligned} \tag{21}$$

For example, with $\lambda_{\text{eff}} = 0.12$, $\gamma_i = \zeta = 1$ and $\delta = \lambda_{\text{eff}}/2$, one finds a rather wide wall with $\Delta_{\text{wall}} \simeq 11.5/T_c \simeq 0.15$ GeV$^{-1}$. For the equation of motion for $\theta$, we define a dimensionless distance $\hat{z} = z/\Delta_{\text{wall}}$ along the bubble wall and obtain

$$\partial_{\hat{z}}^2 \theta + 4(1-g)\partial_{\hat{z}}\theta + B\sin(\theta - \delta_\kappa) + Cg^2(\sin 2\theta)/2 = 0;$$
$$B \equiv 2\kappa \Delta_{\text{wall}}^2; \qquad C \equiv 4h_3 \Delta_{\text{wall}}^2 \rho_c^2 = 64h_3/\lambda_{\text{eff}}. \tag{22}$$

The boundary conditions in the broken and unbroken symmetry phases ($z \gg 0$ and $z \ll 0$ respectively) depend on the values of the parameters $B$ and $C$. By demanding that the derivative of $\theta$ vanishes at $\pm\infty$ it is easy to see that the boundary conditions for the case $B + C < 0$ are

$$\theta(z) = \begin{cases} \delta_\kappa, & z = -\infty \\ \delta_\kappa B/(B+C), & z = +\infty \end{cases} \tag{23}$$

to lowest order in $\delta_\kappa$. However if $B+C > 0$, the boundary conditions are no longer proportional to $\delta_\kappa$. In particular in the case that $\delta_\kappa = 0$, so that CP is explicitly conserved by the Higgs potential, one finds the nontrivial boundary conditions (note that because $B < 0$ this solution *only* exists when $B+C > 0$)

$$\theta(z) = \begin{cases} 0, & z = -\infty \\ \cos^{-1}(-B/C), & z = +\infty \end{cases}. \tag{24}$$

This is an example of spontaneous breaking of CP at finite temperature [9], where CP is conserved at $T = 0$. It is interesting because in this case there will be no constraints from laboratory searches for CP violation on the phase $\theta$. It is straightforward to show that spontaneous CP violation occurs only for sufficiently large values of $\zeta = -\kappa/m_{h^0}^2$,

$$\begin{aligned} \zeta &> \sqrt{F^2 + 3\gamma_A(\delta/\lambda_{\text{eff}})^2} + F; \\ F &\equiv 1.5c/\lambda_{\text{eff}} + 1/4 - 9\delta^2/2\lambda_{\text{eff}}^2 + \sum_{\substack{h^0, A, H \\ +,-}} \gamma_i/16. \end{aligned} \tag{25}$$

using the parameter $c$ defined in (19) and the mass ratio $\gamma_A = m_A^2/m_{h^0}^2$. The constraints (19) and (25) restrict the range of spontaneous CP violation to a rather narrow band in $\zeta$. For example with the particular choice of parameters shown in (34), one finds that it occurs if $7.85 \lesssim \zeta \lesssim 8.70$. However, this is essentially the only constraint there is; one can verify that the upper bound on $\zeta$ in (19) does not conflict with the lower bound (25) unless $\gamma_A$ is larger than the average ratio of the other Higgs masses by the amount

$$\gamma_A - \sum_{i \neq A, h^0} \gamma_i/3 = 8c/\lambda_{\text{eff}} + 5/3 - 8(\delta/\lambda_{\text{eff}})^2. \tag{26}$$

Such a large discrepancy between the masses seems unlikely and would invalidate the use of perturbation theory in the construction of the finite temperature effective potential. We will always assume $\delta_\kappa = 0$ when considering the spontaneous CP violation and focus on the situation $B+C \sim 0$, so that the boundary values of $\theta$ are small and we are justified in treating it as a perturbation.

Since there is no unique choice for the sign of $\theta$ when CP is only spontaneously violated, one would expect that half of all the bubbles that form during the transition produce baryon asymmetries of the opposite sign, which average to zero in the end. However it is conceivable that a small amount of explicit CP violation could be dynamically amplified by the spontaneous effect [23], avoiding the cancellation, so we will keep in mind the possibility.

Let us first however consider the case of explicit CP-violation with $B+C < 0$ and small CP-violating angle $\delta_\kappa \ll \pi$. The linearized equation is

$$\partial_{\hat{z}}^2 \theta + 4(1-g)\partial_{\hat{z}}\theta + (B+Cg^2)\theta + D = 0;$$
$$D \equiv -\delta_\kappa B. \tag{27}$$

Although one might try various techniques for numerically integrating (22), they require specifying boundary conditions at finite values of $z$, where one does not know the exact asymptotic behavior without first having solved the equations. We found that results for the baryon asymmetry due to heavy top quarks were quite sensitive to small uncertainties in the asymptotic behavior, making these methods unsuitable. Instead, it is convenient to rewrite (27) with the kink profile $g$ as the independent variable,

$$4g^2(1-g)^2\partial_g^2\theta + 4g(1-g)(3-4g)\partial_g\theta + (B+Cg^2)\theta + D = 0. \tag{28}$$

The solution can be written as a power series in $g$ or $1-g$, valid on the interval $g \in (0,1)$,

$$\theta(g) = \sum_{k=0}^\infty (\alpha_k + \beta_k g^\gamma)g^k = \sum_{k=0}^\infty (\tilde{\alpha}_k + \tilde{\beta}_k(1-g)^{\tilde{\gamma}})(1-g)^k. \tag{29}$$

The coefficients are given by recursion relations,

$$\beta_k = \frac{(\gamma+k-1)(8(\gamma+k)+12)\beta_{k-1} - (4(\gamma+k-2)(\gamma+k+1)+C)\beta_{k-2}}{4(\gamma+k)(\gamma+k+2)+B};$$
$$\tilde{\beta}_k = \frac{(4(\tilde{\gamma}+k-1)(2(\tilde{\gamma}+k)+1)+2C)\tilde{\beta}_{k-1} - (4(\tilde{\gamma}+k-2)(\tilde{\gamma}+k+1)+C)\tilde{\beta}_{k-2}}{4(\tilde{\gamma}+k)^2+B+C};$$
$$\gamma = (1-B/4)^{1/2} - 1; \qquad \tilde{\gamma} = (-B-C)^{1/2}/2. \tag{30}$$

The analogous relations for $\alpha$ and $\tilde{\alpha}$ are obtained from these by setting $\gamma = \tilde{\gamma} = 0$. $\alpha_k$ and $\tilde{\alpha}_k$ correspond to the inhomogeneous solution and are proportional to D,

$$\alpha_{-1} \equiv 0; \qquad \alpha_0 = -D/B;$$
$$\tilde{\alpha}_{-1} \equiv 0; \qquad \tilde{\alpha}_0 = -D/(B+C). \tag{31}$$

$\beta_k$ and $\tilde{\beta}_k$ correspond to the inhomogeneous solution, so their overall scales are not determined by the equation itself,

$$\begin{aligned} \beta_{-1} &\equiv 0; \quad \beta_0 \text{ undetermined}; \\ \tilde{\beta}_{-1} &\equiv 0; \quad \tilde{\beta}_0 \text{ undetermined}. \end{aligned} \tag{32}$$

However these scales are fixed by demanding that the two series and their derivatives be equal at any intermediate point on the interval. After this, by choosing the appropriate series for the point in question it is always possible to get quite rapid numerical convergence.

For the case of spontaneous CP-violation, where $B + C > 0$, we can again linearize the equation of motion in $\theta$ as long as its value in the broken phase $\theta_0$ is small, that is, if $-B/C$ is not much smaller than unity. Expanding around $\theta_0$ gives an equation similar to to (27) except with a term of the form $Dg^2$ rather than $D$. However one is always free to shift $\theta$ by a constant since this has no effect on the reflection asymmetry of the fermions, and a judicious shift allows us to recover the original form of the equation (27), except that the coefficients must be reparametrized according to

$$\begin{aligned} B &\to B \cos \theta_0; \\ C &\to C \cos 2\theta_0; \\ D &\to B \cos \theta_0 \left( \tan \theta_0 - \frac{1}{2} \tan 2\theta_0 \right), \end{aligned} \tag{33}$$

where $\cos \theta_0 = -B/C$ in terms of the original parameters. After this replacement the solution for $\theta(g)$ is identical to that given above.

In previous papers it has been assumed that the kink ansatz, $\theta(z) = \Delta \theta g(z)$ provides a good approximation to the solution, where $\Delta \theta$ is the difference between the two boundary values. Because an overall additive constant is irrelevant for our imminent goal of determining the asymmetry of fermions reflecting from the domain wall, we can also take the exact solution for $\theta$ to vanish in the symmetric phase. In Figure 1 we compare the profiles of the real solution and the ansatz, using the parameter set

$$\lambda_{\text{eff}} = 0.12, \quad \delta = 0.06, \quad \gamma_i = 1, \quad \kappa \text{ variable}, \tag{34}$$

corresponding to Higgs boson masses of 60 Gev, at the experimental lower limit. For very small $\kappa$ ($\zeta \ll 1$) the solution falls to zero in the symmetric phase ($z = -\infty$) much more slowly than does the tanh ansatz. For intermediate values the two profiles are close to each other, and as $\zeta$ becomes large, the real solution falls to zero before the ansatz does. This behavior continues right up to the critical value (25) of $\zeta$ where spontaneous CP violation begins, at which point increasing $\zeta$ causes the solution to head back toward the tanh ansatz. For $\zeta = 8$ the solution once again lies quite close to the ansatz, only to move again away from it with even larger $\zeta$. The moral of this story is that the tanh ansatz seems, somewhat accidentally, to be a rather good approximation to the solution for certain narrow ranges of $\kappa$ including the natural values $\sqrt{\kappa} \approx m_{h^0}$, but may be poor elsewhere.

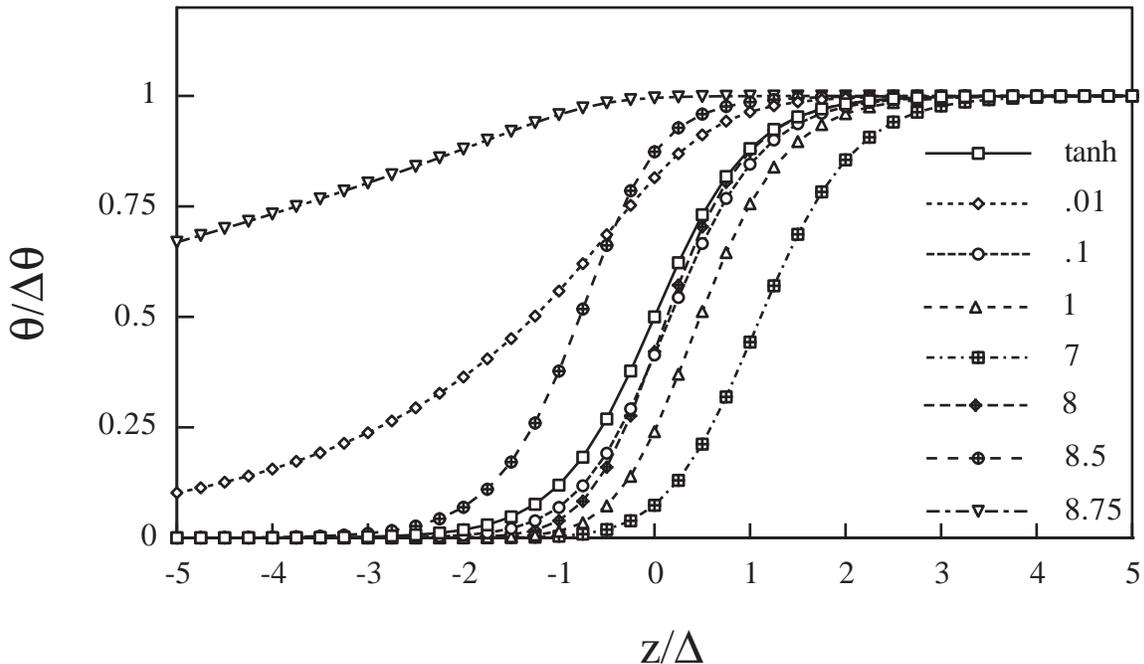

**Figure 1.** The $\theta$-profiles corresponding to the solution of equation (22) for the set of parameters defined in (34) and for varying $\zeta = -\kappa/m_{h^0}^2$. The curves with $\zeta < 7.85$ and $\zeta > 7.85$ correspond, respectively, to explicit and spontaneous breaking of CP symmetry. All curves are scaled by $\Delta\theta$ and shifted so as to be zero at the symmetric phase. The profile corresponding to the tanh ansatz $\theta(z)/\Delta\theta \equiv (1 + \tanh(z/\Delta_{\text{wall}}))/2$ is given by the solid line.

It is interesting to note that the tanh ansatz in fact coincides with the exact solution of $\theta$ for special values of the parameters. This can easily be seen by substituting the guess $\theta = \alpha_0 + \alpha_1 g$ into eq. (28), resulting in the conditions $B = -12$, $C = -16$ in the case of explicit CP violation. These in turn can be solved for $\zeta$ and some linear combination of $\gamma_i$. The solution takes the form

$$\begin{aligned} \zeta &= \frac{-B}{16(a\lambda_{\text{eff}}/\delta^2 - 1)}; \\ \gamma_A &= 2\zeta - C/16. \end{aligned} \quad (35)$$

Normally the first of these equations would be quadratic in $\zeta$ because the parameter $a$ (eq. (6)) depends on $\zeta$ and $\gamma_i$ through eq. (17) for the quartic couplings. However if we assume that all the $\gamma_i$ are equal (with the exception of $\gamma_{h^0} \equiv 1$), the second equation of (35) fixes precisely the same linear combination of $\zeta$ and $\gamma_i$ as appears in $a$. For example, with $\lambda_{\text{eff}} = 0.12$ and $\delta = 0.06$ as above, (35) gives $\zeta = 0.06$ and $\gamma = 1.1$ when $B = -12$, $C = -16$. This explains why the solution for $\zeta = 0.1$ falls so close to the tanh ansatz in figure 1. One can also find that the tanh solution is recovered when $B = -(480)^{1/2}$, $C = 40$ for the case of spontaneous CP violation. These values also correspond to some hypersurface in the physical parameter space, but it is less straightforward to find a representative here than in the case of explicit CP violation above.

We have also found a more analytical solution for $\theta(x)$ which can be expressed as a single integral. This is presented in the Appendix A.

## 4 Fermion Reflection Asymmetry from the Bubble Wall

With the solution for the spatial dependence of the CP-violating phase at hand, we now wish to compute the fermion reflection asymmetry from the bubble wall. We will first consider the scattering to the zeroth order, ignoring the effects of the background to the fermion propagation and then generalize the treatment for the scattering of quasiparticles, i.e. for the effective 1-particle excitations of the plasma at finite temperature. The zeroth order treatment will be adequate for the scattering of fermions with large transverse momentum and in particular for the scattering of the top quark (see appendix B for further details). For light fermions and with small momentum the quasiparticle picture is essential, but even then we will be able to derive the corresponding reflection asymmetries from the zeroth order results by simple mapping of the momentum variable.

## 4.1 Zeroth order equation

The different reflection probabilities for fermions and antifermions arise from having a spatially varying, complex mass in the Dirac equation,

$$(i\slashed{\partial} - m(z)P_R - m^*(z)P_L)\Psi(t,z) = 0, \tag{36}$$

where $P_L$ and $P_R$ are the chirality projection operators. The mass is given by replacing the Higgs fields in eq. (1) by their vacuum expectation values,

$$m(z) = -y_2 \langle \Phi_2 \rangle = -\frac{y_2}{\sqrt{2}}\rho(z)e^{-i\theta(z)/2}. \tag{37}$$

The fermion states that interact with the bubble wall are eigenstates of energy, not momentum, so one takes $\Psi(t,z) = e^{-iEt}\psi(z)$. In the chiral representation of the Dirac matrices, the Dirac equation then separates into two equations for a pair of two-component spinors which we shall call $\psi_1 = (L_-, R_+)^T$ and $\psi_2 = (R_-, L_+)^T$, where the letter denotes the chirality and the subscript the direction of motion:

$$i\partial_z \psi_1 = \begin{pmatrix} E & -m \\ m^* & -E \end{pmatrix} \psi_1; \qquad i\partial_z \psi_2 = \begin{pmatrix} E & -m^* \\ m & -E \end{pmatrix} \psi_2. \tag{38}$$

The two equations are thus identical except for the interchange $m \leftrightarrow m^*$. The boundary conditions for $\psi_1$, describing an incoming wave from the symmetric phase ($z < 0$) plus reflected and transmitted parts, are

$$\begin{aligned}
\psi_1(z) &= \begin{pmatrix} R(p)e^{-ipz} \\ e^{ipz} \end{pmatrix}, & z \ll 0; \\
\psi_1(z) &= \frac{T(p)}{\sqrt{2p'(p'+E)}} \begin{pmatrix} m^* \\ E+p' \end{pmatrix} e^{ip'z}, & z \gg 0,
\end{aligned} \tag{39}$$

where $R(p)$ and $T(p)$ are the reflection and transmission amplitudes, $m$ is the asymptotic value of $m(z)$ deep in the broken phase, and $p' = \sqrt{E^2 - |m|^2}$. We use the prime to distinguish $p'$ from the momentum deep in the symmetric phase, $p = E$. The boundary conditions are satisfied only for certain values of $R(p)$ and $T(p)$, for which we want to solve. The boundary conditions for $\psi_2$ are the same except for the replacements $m^* \to m$, $R \to \bar{R}$ and $T \to \bar{T}$. Then the asymmetry in the reflection probabilities for $L_- \to R_+$ and $R_- \to L_+$ is

$$\Delta R(p) \equiv |R(p)|^2 - |\bar{R}(p)|^2. \tag{40}$$

The whole analysis can be repeated for the antiparticles simply by letting $E \to -E$ in the original Dirac equation. It is straightforward to show that the respective equations for $\bar{\psi}_1 = (\bar{R}_-, \bar{L}_+)^T$ and $\bar{\psi}_2 = (\bar{L}_-, \bar{R}_+)^T$ are the same as those for $\psi_1$ and $\psi_2$, except for the change $m \to -m$. But the overall sign of the mass can have no effect on measurable quantities, so it follows that the reflection probabilities for $\bar{\psi}_1$ and $\bar{\psi}_2$ are also $|R(p)|^2$ and $|\bar{R}(p)|^2$, respectively.

For particles incident from the broken phase, we can easily find that the reflection probabilities are related to those of the above situation by

$$\begin{aligned} |R^b_{R \to L}|^2 &= |R^s_{L \to R}|^2; \\ |R^b_{L \to R}|^2 &= |R^s_{R \to L}|^2, \end{aligned} \tag{41}$$

where we have indicated the particle chiralities in the subscripts and the superscripts $b$ and $s$ refer to the broken and symmetric phases. The relations can be derived starting from the solution for $\psi_1$ obtained above (which describes the process $L \to R$ in the symmetric phase) and taking its charge conjugate, $\psi'_1 = \sigma_2 \psi_1^*$. This is a time reversal of the original solution, and it satisfies the same equation as $\psi_1$ except for the change $m \to -m$. Therefore if we form a third spinor $\psi''_1$ made in just the same way as $\psi'_1$ except starting from the Dirac equation with $m \to -m$, it will be a solution to the original equation (38) for $\psi_1$. If we denote the reflection and transmission amplitudes for $\psi_1$ as $R(m)$ and $T(m)$ respectively, then those of $\psi''_1$ are $R^*(-m)$ and $T^*(-m)$. Next we can form a linear combination of $\psi_1$ and $\psi''_1$ (remember that $\psi''$ goes backward in time),

$$(R^*(-m)\psi_1 - \psi''_1)/T^*(-m) \tag{42}$$

chosen so as to exactly cancel the incoming wave from the symmetric phase and to normalize the incoming wave from the broken phase to unity. Thus our new solution describes reflection $R \to L$ of a particle incident from the broken phase, and the reflection coefficient is seen to be $-R^*(-m)T(m)/T^*(-m)$. When we square this and use the fact that no observable can depend on the sign of the mass, we immediately get the first of eqs. (41). This result is a consequence of CPT invariance.

The reflection amplitudes can be numerically computed from (38) and (39) by a straightforward shooting algorithm to integrate the two differential equations. We have used this method to verify the results of previous authors. However it breaks down when the quark mass in the broken phase starts to greatly exceed the inverse wall width, that is when

$$\xi \equiv m_c \Delta_{\text{wall}} \equiv \frac{4m_0}{m_{h^0}} \gg 1. \tag{43}$$

(where $m_c$ and $m_0$ are the fermion mass values at $T = T_c$ and $T = 0$, respectively). This is apparently because the solution begins to undergo many oscillations over the region of the bubble wall, which makes it prohibitively difficult to numerically solve the equation between the two asymptotic regimes outside the wall. This is also the regime where the overall reflection coefficients are exponentially suppressed in the fermion mass, so that this regime should make a subdominant contribution to the total baryon asymmetry. We have used the perturbative method of Funakubo *et al.* [24] to compute $\Delta R$ in this case. They found that for the lowest order in $\theta \ll \pi$ and for the particular wall profile $g(z)$ given by (41), the asymmetry is given by an integral involving $\theta(z)$, $g(z)$, and the unperturbed negative- and positive-chirality wave functions $\phi_p^\pm(z)$ of the fermions,

$$\Delta R(p) = C_p \int_{-\infty}^{\infty} dz\, \phi_p^+(z) \phi_p^-(z) \frac{d}{dz}(g(z)\theta(z)) + c.c. \tag{44}$$

The complex constant $C_p$ in front of integral can be expressed in terms of ratios of gamma functions and the wave functions $\phi_p^\pm(z)$ are expressible in terms of hypergeometric functions [24].

Some typical profiles for the reflection asymmetry as a function of momentum are shown in figure 2. For later purposes it turns out that a simple exponential provides a reasonably good fit for any value of the mass,

$$\frac{\Delta R(p_z)}{\Delta \theta} \cong \begin{cases} A(\xi)e^{-(p_z - m)/w(\xi)}, & p_z > m \\ 0, & p_z < m \end{cases} \tag{45}$$

with a height and width that depend on the mass. $\Delta R$ vanishes for $p_z < m$ because then both particles and antiparticles are totally reflected by the bubble wall. The actual functions $\Delta R(p_z)$ go smoothly to zero as $p_z \to m$, as shown in figure 2, but this occurs on a scale much shorter than the width $w(\xi)$, so that the exponential is not a bad approximation. Little error is made by using eq. (45) in later expressions for the baryon asymmetry since these are momentum integrals which do not especially weight the threshold region.

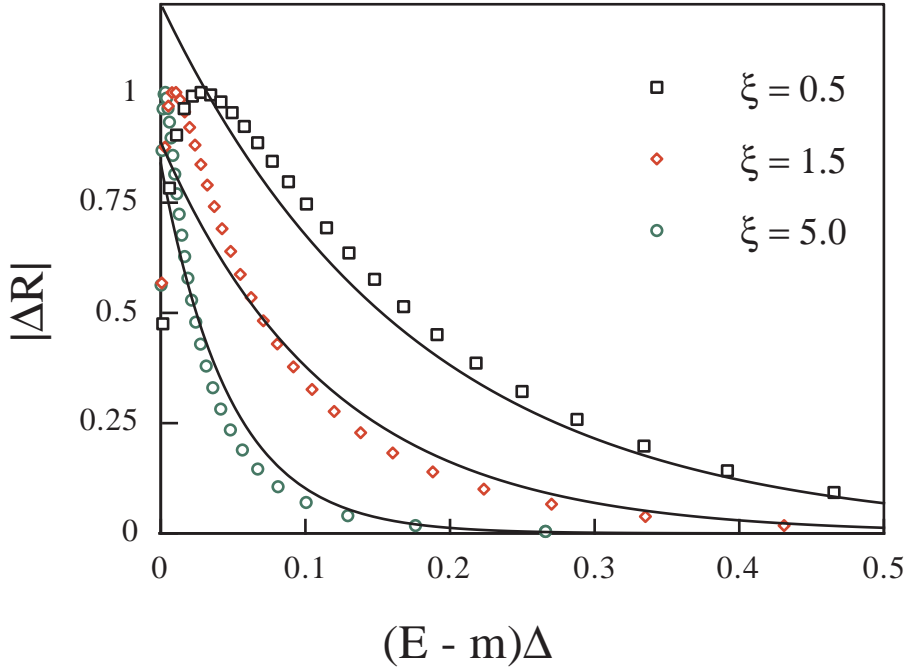

**Figure 2.** The scaled $\Delta R(p)/\Delta\theta$ profiles defined by eqs. (38−40) for certain representative values of the mass parameter $\xi = m(T_c)\Delta_{\rm wall}$. The solid lines are the exponential fits given by eq. (45).

A curious property of eq. (44) is that when one uses the tanh ansatz $\theta(z) \sim g(z)$, the sign of the asymmetry oscillates as a function of the quark mass, changing at small integer values of the inverse wall width, $\xi = 2, 3, 4$. Thereafter it falls to a value smaller than our computational accuracy. The real solutions for $\theta(z)$ typically display no such behavior, and $\Delta R$ falls much more slowly with increasing fermion mass than for the ansatz. Using our complete numerical code we have verified that this is truly a behaviour of the ansatz in the small $\theta$ limit and not an artifact of the linear approximation leading to (44). However, even for the ansatz, if the change in $\theta$ is sufficiently large that (44) is no longer valid, the full numerical solution of the Dirac equation reveals no such oscillatory behavior. The difference is shown in figure 3, where we plot the maximum value of $|\Delta R(p_z)|$ versus $\xi$ for the first two cases.

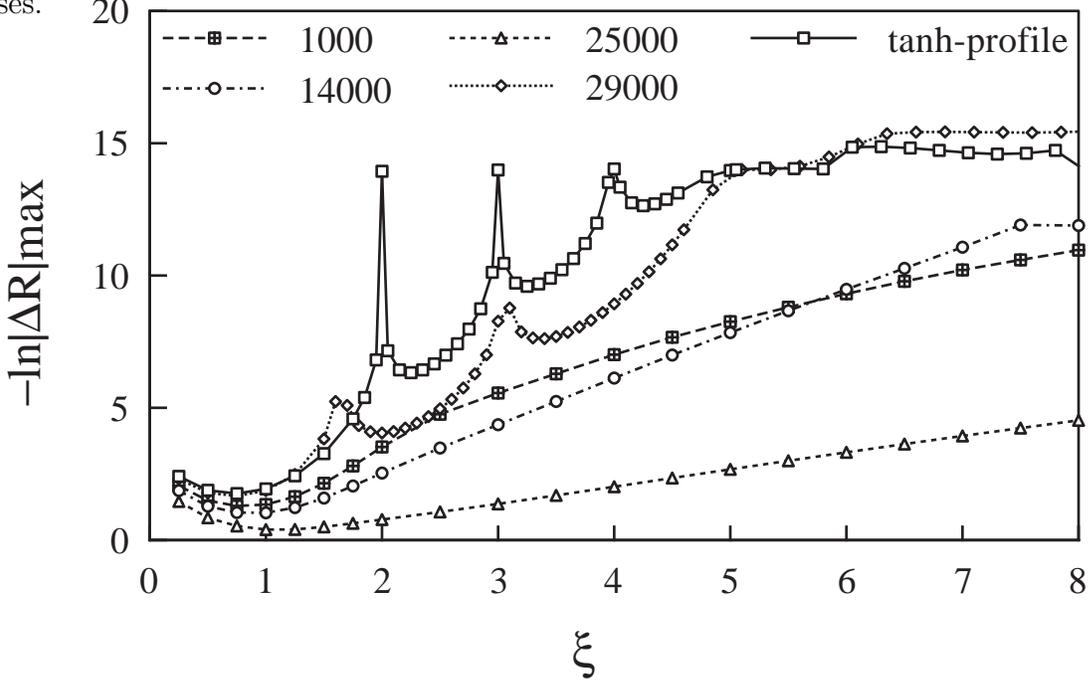

**Figure 3.** The dependence of the maximum value of the $\Delta R$ profile on the mass parameter $\xi$, for different values of the Higgs potential parameter $\kappa$ in units of GeV$^2$ (assuming a Higgs boson mass of $m_{h^0} = 60$ GeV.) For comparison with figure 1, the values of $\zeta = -\kappa/m_{h^0}^2$ corresponding to these $-\kappa$ values are 0.28, 3.9, 7 and 8, respectively.

We have also examined how the width $w(\xi)$ of the $\Delta R$ profiles varies as a function of the fermion mass. We have defined it to be the area under the curve $\Delta R(p_z)$ divided by the maximum value of $\Delta R$ discussed above. The dependence is shown for typical values of the model parameters in figure 4. In fact we find that $w(\xi)$ is largely independent of the potential parameter $\kappa$. For the region of fermion masses shown in figure 4, it is fit well by the expressions

$$\begin{aligned} w(\xi)/m &= -(1.1\ln\xi + 0.54), & \xi &< 0.3; \\ w(\xi)/m &= 0.19\,\xi^{-1.2}, & 0.3 &< \xi < 0.7; \\ w(\xi)/m &= 0.15\,\xi^{-1.8}, & \xi &> 0.7; \end{aligned} \tag{46}$$

where henceforth $m$ stands for the mass of the fermion at the critical temperature. The smaller values of $\xi$ are of interest for the tau lepton.

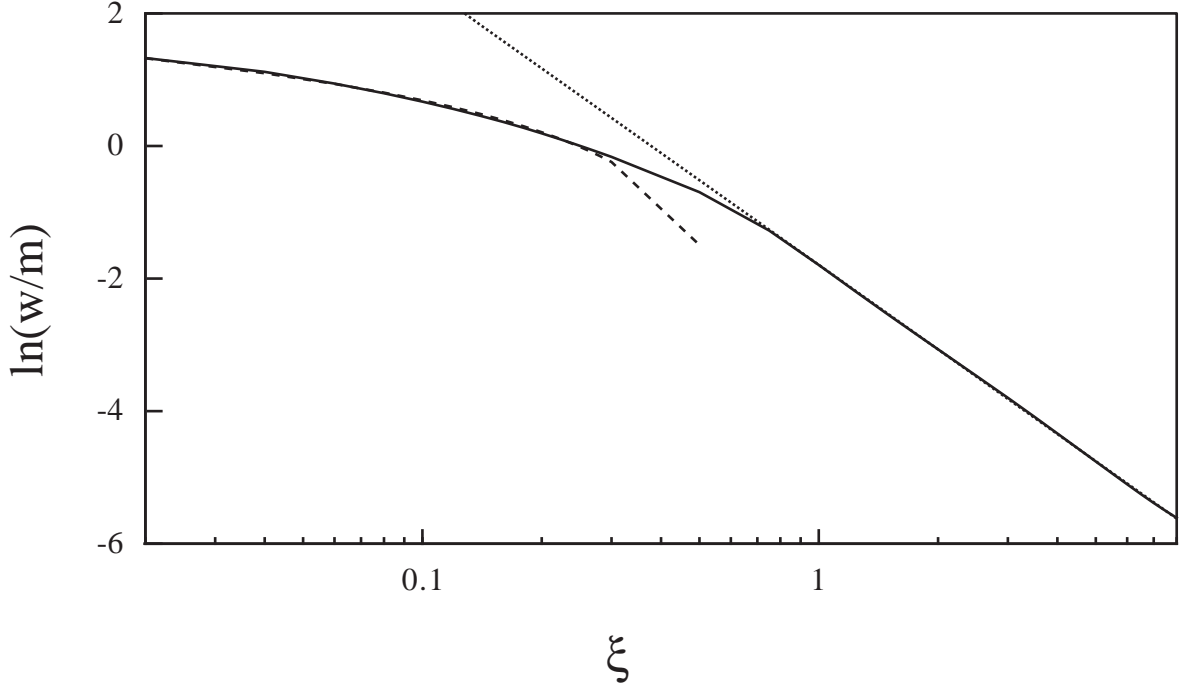

**Figure 4.** The width in momentum space $w(\xi)$ of our fit (45) for the asymmetry in the reflection probabilities $\Delta R(p)$, along with our fits (46) to $w(\xi)$ for large and small $\xi = m\Delta_{\text{wall}}$.

To make contact with the physical situation of interest, we note that for the choices of parameters we have been using for the Higgs potential (only the value $\lambda_{\text{eff}} = 0.12$ is relevant here), the dimensionless quantity $\xi = m\Delta_{\text{wall}}$ that characterizes the fermion mass turns out to be

$$\begin{aligned} \xi &= 11.7, \quad \text{top quark} \\ \xi &= 0.33, \quad \text{bottom quark} \\ \xi &= 0.12, \quad \text{tau lepton} \end{aligned} \qquad (47)$$

For the top quark this means that the reflection coefficient is extremely small except in a very narrow region of momentum space. We find that the height times width of the reflection asymmetry profile is of order $Aw \sim 10^{-10}m$, which is seven orders of magnitude smaller than that of the bottom quark. Henceforth we will ignore the top quark contribution to the process of forming the baryon asymmetry.

For future reference we tabulate certain values of $w(\xi)$, along with the corresponding values of $A(\xi) \equiv |\Delta R|_{\max}$, using the parameters of eq. (34) and the representative value $\kappa = -1000$:

$$\begin{aligned}
\xi &= 0.06; & w/m &\simeq 2.6; & A &\simeq e^{-3.2} \\
\xi &= 0.12; & w/m &\simeq 1.8; & A &\simeq e^{-2.7} \\
\xi &= 0.16; & w/m &\simeq 1.3; & A &\simeq e^{-2.5} \\
\xi &= 0.33; & w/m &\simeq 0.7; & A &\simeq e^{-1.9}
\end{aligned} \qquad (48)$$

## 4.2 Thermal corrections: small momentum regime

The propagation of fermions is affected by the ambient high temperature plasma in the early Universe. Therefore the Dirac equation used in the previous subsection, which assumed that the fermions obey the usual vacuum dispersion relations, apparently needs some modification. We give a more detailed derivation of the modified Dirac equation in the appendix B, where we argue that the zeroth order treatment given above is adequate to use in the large transverse momentum region, and also derive the equations used below to study the small momentum limit, where the effects of the background are most important.

Whenever the fermion momentum is small compared to the thermal masses, denoted by $\omega_L$ and $\omega_R$ for the two chiralities of a given fermion species, induced by interactions with the plasma [25, 26], the dispersion relations for the $i$th chirality, in the rest frame of the plasma, are changed to [26] (see also appendix B):

$$\begin{aligned}
\omega &= \omega_i \pm |\vec{k}|/3, & &\text{symmetric phase;} \\
\omega &= \omega_0 \pm ((\Delta_\omega/2 \pm |\vec{k}|/3)^2 + |m|^2/4)^{1/2}, & &\text{broken phase,}
\end{aligned} \qquad (49)$$

using the definitions

$$\omega_0 = (\omega_L + \omega_R)/2 \quad \text{and} \quad \Delta_\omega = \omega_L - \omega_R. \qquad (50)$$

They are shown in figure 5. The modes with the negative slope $d\omega/dk$ close to the origin are called holes [27], or abnormal [26], because their group velocity is opposite to their momentum, and since they do not exist at low temperatures.

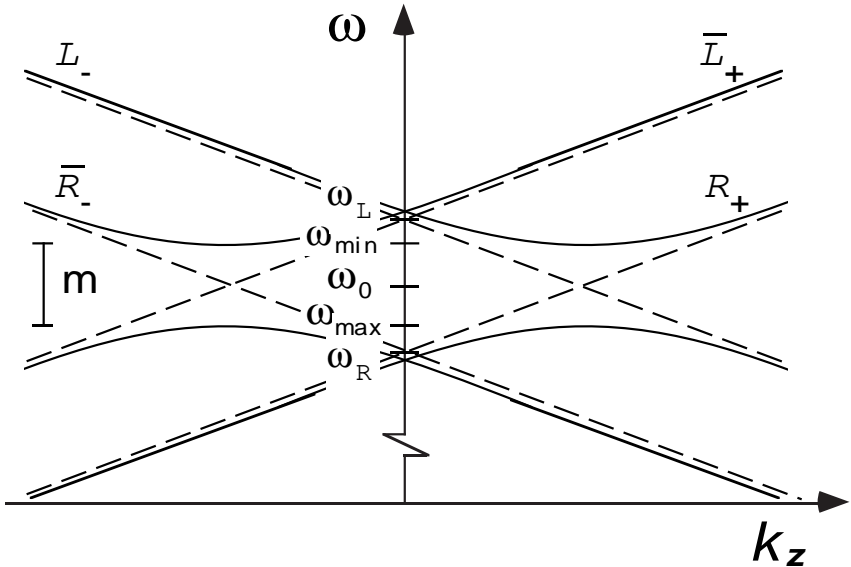

**Figure 5.** Schematic view of the fermion dispersion relations in the small momentum limit corresponding to equations (49). The solid lines pertain to the broken and the dashed lines to the symmetric phase. The mass gap in the former, near the energy $\omega_0$, has half of its zero-temperature value.

When the momentum is purely in the $z$ direction, the dispersion relations correspond to an effective Lagrangian whose resulting Dirac equation is, instead of (38),

$$\frac{i}{3}\partial_z \psi_1 = \begin{pmatrix} \omega - \omega_L & -m/2 \\ m^*/2 & -\omega + \omega_R \end{pmatrix} \psi_1; \quad \frac{i}{3}\partial_z \bar{\psi}_2 = \begin{pmatrix} \omega - \omega_R & -m^*/2 \\ m/2 & -\omega + \omega_L \end{pmatrix} \psi_2. \quad (51)$$

For the antiparticles, one must change not only the sign of $\omega$ but also the $\omega_i$'s since purely thermal effects increase the energies of both particles and antiparticles equally.

The quasiparticle dispersion relations given here are a linear approximation valid only for momenta much smaller than the scale $\omega_0$. In particular the abnormal mode energies do not really fall monotonically to zero but reach a minimum and start to rise again. However this happens on a scale $\omega_0$, which is much larger than the scale of the exponential decay in $\Delta R(p)$ apparent in our previous solutions of (38), and which will also be supplied by the solutions of (51) *a posteriori*. Therefore, setting aside the possible dependence on the parallel momentum, the linear approximation is reasonable for computing the effects of the quasiparticles on baryon production.

Despite the differences between (51) and the original Dirac equation (38), there is a simple mapping between their reflection coefficients because each equation can be transformed into the other. Let $\tilde{z} = 3z$, $\psi_{1,2} = e^{\pm i \Delta_\omega \tilde{z}/2} \widetilde{\psi}_{1,2}$, and $\tilde{\omega} = \omega - \omega_0$. It is straightforward to show that the new equations resulting from this transformation are the same as one would obtain from the Dirac equation (38) by making the replacement $E \to \tilde{\omega}$ and $m \to m/2$ in the latter. The reflection asymmetry is given by exactly the same function $\Delta R(p)$, but the correspondence between $p$ and the actual (outgoing) momentum $k$ is

$$p = k/3 + \begin{cases} \Delta_\omega/2, & L_+, \; R_- \text{ modes} \\ -\Delta_\omega/2, & R_+, \; L_- \text{ modes} \end{cases} . \tag{52}$$

This applies to the excitations, whose energies exceed $\omega_0$. The Dirac equations for the modes with $\omega < \omega_0$ are the same except for the opposite sign of $\tilde{\omega} = \omega - \omega_0$. The sign change means that there is a mapping between the Dirac equations for $\omega > \omega_0$ and $\omega < \omega_0$ modes which makes them look the same except for the interchange of $m$ and $m^*$. Therefore the reflection asymmetry changes sign along with $\omega - \omega_0$ [28]. To be explicit, recalling that $\Delta R(p)$ is the difference in reflection probabilities between the processes $L_- \to R_+$ and $R_- \to L_+$ for $\omega > \omega_0$, $-\Delta R(p)$ is the reflection asymmetry for the modes with $\omega < \omega_0$, which are depicted in figure 5.

We assumed that the components of momentum parallel to the wall were zero in this discussion. For the usual dispersion relations which are valid at large momenta this is no limitation because one can always boost to the frame where these components are zero to solve the Dirac equation. In the small momentum region however, the dispersion relations of the quasiparticles are clearly not Lorentz invariant, so that if one does the same boost, they take on a different form which is incompatible with the boundary conditions of the Dirac equation as we have written them. The exact treatment would be quite cumbersome, so we will compromise by ignoring the dependence on the parallel momenta when they are smaller than $\omega_0$. When they are greater than $\omega_0$, the usual dispersion relations become appropriate and we can use the first form (38) of the Dirac equation.

## 4.3 Decoherence during the reflection

We have so far completely ignored the effects of damping in our treatment of the reflection from the wall. In reference [29] it was argued that the continuous scatterings off the background particles experienced by the reflecting quarks leads to a significant loss of coherence of the wave function. They accounted for the scatterings by including the complex part of the quark self energy into the Dirac equation, and found a tremendous suppression of the final baryon asymmetry in the minimal standard model. The issue of how to correctly account for the decoherence phenomenon is still controversial [30], and we do not wish to get into the details of that argument here. Here we will demonstrate that the neglect of damping is much better justified for the situation in which we are interested.

Let us begin with the damping in the small-momentum region. The damping rate of a fermion at zero momentum was first computed in a gauge invariant way in references [31, 32] with the result:

$$\gamma_F(k=0) = a(N, N_f) C_F(N) \frac{g^2 T}{16\pi}, \tag{53}$$

where $C_F = (N^2-1)/2N$ is the usual casimir operator eigenvalue and the constant $a(N, N_f)$ has a weak dependence on the group index $N$ and the number of fermion families $N_f$. In QCD with three families, references [31, 32] give $a(3,3) \simeq 5.7$ which, with $\alpha_s \sim 0.1$, leads to a rate $2\gamma_q \sim 0.19T$. This result was used by ref. [29] to obtain a mean free path of the quarks of $\ell_q \sim 1/6\gamma_q \sim 0.9/T$, which is significantly less than the smallest expected wall widths $\Delta_{\text{wall}} \gtrsim \text{few}/T$.

However, this large result for the quark damping rate is almost exclusively due to strong interactions. There exists no standard model computation of $\gamma$ for the leptons in the literature, but from the results of [32] for pure $SU(2)$, $a(2,2) \simeq 5.8$ and $a(2,4) \simeq 6.3$, one can estimate that $a_\tau \simeq 6$. Then (53) straightforwardly gives $\gamma_\tau \simeq 0.04T$, hence $\ell_\tau \sim 4.4/T$. This result is comparable to the wall width predicted in the model under consideration, so we expect that scattering of low-momentum leptons by the plasma will not strongly damp their quantum mechanical reflection from the wall, even if the more restrictive picture of the ref. [29] is the correct one.

At large momenta there are no problems with the infrared properties of the gauge interactions that led to the difficulties in the evaluation of $\gamma$ at low momenta [31, 32] and the damping rate is given by the usual scattering computation. One then expects that the damping length would be roughly one third of the diffusion length. Thus for quarks we estimate that $\ell_q \sim D_q/3 \simeq 2/T$ and for leptons $\ell_L \sim 40/T$. One might be led to believe that at least for leptons, the neglect of damping during the reflection process is a good approximation, given our determination of $10/T$ for the wall width. However one must keep in mind that the important reflections are coming from particles with small momenta perpendicular to the wall, whose other momentum components are typically of order $T$, so that they approach the wall at a glancing angle and therefore typically undergo several interactions during their traversal of the wall [8]. In our treatment below we will distinguish between the particles with large and small momenta parallel to the wall, the latter of which are much less sensitive to the decoherence effects.

To conclude this section we note that it is not yet clear how to consistently compute the effects of decoherence in the present mechanism of baryogenesis. Nevertheless, in the end we will find that the reflection of tau leptons dominates baryon production and moreover, for these particles the region of phase space where all components of the momenta are small (so that they are approaching the wall from a sharp angle and do not undergo many scatterings) dominates over the large momentum region. Therefore we do not expect that the decoherence effects will be crucial for our final results.

## 5   Fermion Transport

To compute the baryon asymmetry resulting from fermions reflecting off the wall, it is necessary to understand the diffusion of reflected fermions back into the symmetric phase. There exist several different ways of treating this in the literature: Monte Carlo simulations [7], the diffusion equation [8], and solving a more exact form of the Boltzmann equation, called the Fokker-Planck equation [33]. Although the Fokker-Planck equation should in principle be more accurate, more work is needed to establish why it gives different results from the diffusion equation, which has heretofore received more attention. For ease of comparison with previous work, we will also adopt the diffusion equation as our framework for solving the transport problem.

Our approach will proceed in three steps. First we compute the flux of net right-handed and left-handed fermion numbers into the symmetric phase in the vicinity of the bubble wall. This results in an initial chiral asymmetry in front of the wall, in which we are interested because it is what biases sphalerons to produce baryons, since the sphaleron rate is proportional to the asymmetry in left-handed fermions. However, flavor-changing processes will redistribute the initial asymmetry amongst the various species of particles, so we must next take into account the interactions which are fast compared to the diffusion rate, and find new initial values for the densities near the wall after chemical equilibrium is established. In the third step we consider the diffusion of the particle asymmetries into the symmetric phase, incorporating the Debye screening of hypercharge using the results of reference [37]. The goal of these computations is a spatial profile for the total left-handed fermion number which can be used to compute the rate of baryon number violation, assumed to be the slowest process of interest in the problem. Integration of this rate finally gives the baryon asymmetry.

## 5.1 Fermion Flux in the Symmetric Phase

The first step toward generating a baryon asymmetry is to create an asymmetry in the density of left-handed particles in the symmetric phase, since it is these which drive sphalerons to create baryons. The left-handed asymmetry arises due to four contributions: left-moving, right-handed particles $R_-$ reflect into right-moving, left-handed particles $L_+$ with reflection probability $|R(p)|^2$, where $p$ labels the momentum in the symmetric phase; using eq. (41), $L_+$ fermions are transmitted from the broken phase with probability $1-|R(p)|^2$; the analogous processes with antiparticles give a cancelling contribution, however with a different probability $|\bar{R}(p)|^2$ because of the CP violation in the wall. For a given momentum, this gives a left-handed current of

$$J_L \propto |R|^2 f_s(R_-) + (1-|R|^2) f_b(L_+) - |\bar{R}|^2 f_s(\bar{R}_-) - (1-|\bar{R}|^2) f_b(\bar{L}_+), \tag{54}$$

where $f_{s,b}(X)$ denotes the Fermi-Dirac distribution function for species $X$ with subscript $s$ or $b$ showing whether the corresponding particle is propagating in the symmetric or the broken phase. Because the wall is moving, $f(X)$ will be different for left- and right-movers. It is however the same for particles and antiparticles, so that using eq. (40), (54) becomes

$$J_L \propto \Delta R(p)(f_s(R_-) - f_b(L_+)). \tag{55}$$

Thus the asymmetry vanishes in the absence of either CP violation or the velocity of the wall, as expected. The current of right-handed fermions has the opposite sign due to the CPT theorem (as can be seen from comparing the Dirac equations for right- and left-handed particles),

$$J_R \propto -\Delta R(p)(f_s(L_-) - f_b(R_+)). \tag{56}$$

To the extent that left- and right-handed particles have identical dispersion relations, hence equal distribution functions, the sum of $J_L$ and $J_R$ is zero, so that there is no net current of baryon or lepton number. In fact the two chiralities get different thermal energy shifts due to their different interactions with the background plasma particles; this will be relevant when considering the contributions to the baryon asymmetry from the reflected quarks.

The full expression for the flux is an integral over all the momenta of the expressions like those above, weighted by the group velocity of the right-moving particles which is obtained from their dispersion relations. We will divide the momentum space into two regions, depending on whether the momenta are large or small compared to the thermal self-energies. The large-momentum region gives a contribution to the flux which is similar to what we would have computed with the usual dispersion relations, because in this region the temperature corrections become small and the Lorentz symmetry is approximately restored. The small-momentum region gives new contributions associated with the normal and abnormal quasiparticle excitations (49).

### 5.1.1 Large momentum region

We will first concentrate on the large-momentum region. For the flux of left-handed particles per color degree of freedom ($N_c = 3$ for quarks, 1 for leptons), this region contributes

$$\frac{J_L^{\text{l.m.}}}{N_c} \simeq \int_{|m|}^{\infty} \frac{dk_z}{2\pi} \Delta R(k_z; |m|) \int_{\omega_0}^{\infty} \frac{dk_{||} k_{||}}{2\pi} \frac{k_z}{E_1} \left( f(E_1 - vk_z) - f(E_1 + vk'_z) \right) \left( 1 - f(E_1 + vk_z) \right), \tag{57}$$

where the factor $k_z/E$ is the group velocity in the $z$-direction and

$$\begin{aligned}
E_1 &= \sqrt{|\vec{k}|^2 + |m|^2 + 2\omega_L^2}, \\
k'_z &= \sqrt{k_z^2 - |m|^2}.
\end{aligned} \tag{58}$$

Note that the lower limit on the parallel momentum integration in (57) has been somewhat arbitrarily chosen to be $k_{||,\text{min}} = \omega_0$; we will come back to this shortly. To a good approximation the $k_z$ dependence of $E_1$ can be neglected in the phase space distribution. Then taking the limit of small wall velocity we can expand the phase space functions in $v$ after which the $k_{||}$ integral is elementary. Finally, using the form (45) for the reflection asymmetry $\Delta R$, we obtain the result

$$J_L^{\text{l.m.}}/N_c \simeq \frac{v \Delta \theta A w |m|^2}{4\pi^2} r_L(w/|m|) \, f(E_1^L)(1 - f(E_1^L)/2), \tag{59}$$

where

$$\begin{aligned}
E_1^L &= \sqrt{\omega_0^2 + 2\omega_L^2 + |m|^2}; \\
r_L(w/|m|) &= \left( 1 + 2w/|m| + 2w^2/|m|^2 + K_2(|m|/w) e^{|m|/w} \right).
\end{aligned} \tag{60}$$

Here $K_2(x)$ is the Bessel function of the second kind. For small $w/|m|$ the function $r_L(w/|m|)$ approaches unity, which is the appropriate limit to take for the top quark. For all the other fermions $w/|m|$ is not small and one obtains $r_L(w/|m|) \simeq (2 + 4w/|m| + 4w^2/|m|^2)$. For example for the tau lepton with $\xi \simeq 0.12$, $w/|m| \simeq 1.8$ and $r_L \simeq 22$, while for the bottom quark with $\xi \simeq 0.33$, $w/|m| \simeq 0.7$ and $r_L \simeq 6.2$. Thus one sees that the flux is rather sensitive to the fermion mass.

The energy $E_1^L$ reflects how the flux depends on the choice $\omega_0$ for the lower limit of $k_{||}$-integration. This value was chosen to insure that the finite temperature corrections are small at higher momenta, but we might have instead used $k_{\|,\mathrm{min}} = \mathcal{O}(\mathrm{few})\omega_0$. In order to see how this ambiguity affects our results, we need to know the thermal masses of the bottom quark and the tau lepton (the top quark will no longer concern us because its reflection coefficient is so small that it makes a negligible contribution compared to these lighter particles); at the one-loop level they are given by

$$\begin{aligned}
\omega_{bL}^2/T^2 &= g_s^2/6 + 3g^2/32 + g'^2/288 + y^2/16 \simeq 0.40 \\
\omega_{bR}^2/T^2 &= g_s^2/6 + g'^2/72 \simeq 0.24 \\
\omega_{\tau L}^2/T^2 &= (3g^2 + g'^2)/32 \simeq 0.044 \\
\omega_{\tau R}^2/T^2 &= g'^2/8 \simeq 0.016.
\end{aligned} \quad (61)$$

We have ignored all Yukawa couplings except for that of the top quark, $y = 1.4$ and we evaluated the gauge couplings at $M_Z$ (which is close to the critical temperature in our model): $\alpha_s = 0.12$, $g^2 = 0.42$ and $g'^2 = 0.13$. Using these numbers one finds that $\beta E_{1\tau}^L \simeq 0.34 - 0.5$ and correspondingly $f(E_{1\tau}^L) \simeq 0.42 - 0.38$ when the lower limit is varied over the range $k_{\|,\mathrm{min}}^2 = (1-5)\omega_0$. Similarly for the bottom quark we find $\beta E_{1b}^L \simeq 1.1 - 1.5$ and $f(E_{1b}^L) \simeq 0.26 - 0.18$. To a reasonably good accuracy then the contributions to the left-handed flux coming from the high-momentum region are, using (48),

$$\begin{aligned}
J_L^{\mathrm{l.m.}}(\tau) &\simeq 2 \times 10^{-2} v\, \Delta\theta\, m_\tau^3 \\
J_L^{\mathrm{l.m.}}(b) &\simeq 9 \times 10^{-3} v\, \Delta\theta\, m_b^3.
\end{aligned} \quad (62)$$

Thus the initial asymmetry in left-handed $(B+L)$ due to bottom quarks from the high-momentum region is larger by a factor of 16 than that due to tau leptons. (Remember that each quark carries $B = 1/3$.)

### 5.1.2 Small momentum region

We now turn to the contribution to the flux coming from the small-momentum region. This region of phase space was found to be crucial for standard model baryogenesis [26], because only at small momentum does one have the hope of avoiding GIM suppression in the CP-violation arising from the CKM-matrix. At first it would appear that this region does not have the same significance in the present mechanism, where the CP-violation comes from the complex phase of the Higgs field, but we will find that it is actually more important than the large-momentum regime. We first compute the reflected current in one spatial dimension (1D) model and then estimate the 3D-current from that of 1D using a simple phase space argument.

After some straightforward algebra one can show that the 1D left-handed particle flux, for example, is given in the small momentum region by

$$\begin{aligned} J_L^{\text{s.m.}}/N_c &= \int_{\omega_{\min}}^{\infty} \frac{\mathrm{d}\omega}{4\pi} \Delta R(\omega - \omega_0; \tfrac{|m|}{2})(1 - f(\omega + vk_L^N)) \left\{ f(\omega + vk_R^N) - f(\omega + vk_L^{N'}) \right\} \\ &- \int_{\omega_{co}}^{\omega_{\max}} \frac{\mathrm{d}\omega}{4\pi} \Delta R(\omega_0 - \omega; \tfrac{|m|}{2})(1 - f(\omega + vk_L^A)) \left\{ f(\omega + vk_R^A) - f(\omega + vk_L^{A'}) \right\} \end{aligned} \quad (63)$$

where the limits of integration are

$$\begin{aligned} \omega_{\min} &= \omega_0 + |m|/2; \\ \omega_{\max} &= \omega_0 - |m|/2, \end{aligned} \quad (64)$$

and the momenta the left-handed particle transmitted into the symmetric phase, the right-handed particle incident from the symmetric phase and the left-handed particle transmitted from the broken phase, for the normal and abnormal modes, are respectively

$$\begin{aligned} k_L^N &= k_L^A = 3(\omega - \omega_L); \\ k_R^N &= k_R^A = 3(\omega_R - \omega); \\ k_L^{N'}, k_L^{A'} &= 3 \left( -\frac{\Delta_\omega}{2} \pm \sqrt{(\omega - \omega_0)^2 - \frac{|m|^2}{4}} \right). \end{aligned} \quad (65)$$

In arriving at (63) we have transformed the momentum integral into an integral over energy using the relation $\mathrm{d}k(\mathrm{d}\omega/\mathrm{d}k) = \mathrm{d}\omega$, where $(\mathrm{d}\omega/\mathrm{d}k)$ is the group velocity of the excitation, and expressed the relevant momentum variables (52) in terms of the energy difference $\omega - \omega_0$. The relative minus sign between the two terms was explained below eq. (52). Additional factor of $1/2$ included to the phase space measure comes from the wave function renormalization of the incoming flux (Appendix B) and $1 - f(\omega + vk_L^{N,A})$ is the Pauli blocking factor. The integration limits (64) have a simple interpretation, as illustrated in fig. 5: for any energy in between $\omega_{\max}$ and $\omega_{\min}$ the states are totally reflected because they do not have enough energy to penetrate the broken phase. The cutoff $\omega_{co}$ is due to the fact that the linear dispersion relations break down and the quasiparticles become unstable at the momenta that would correspond to such small energies. However the actual value of $\omega_{\rm co}$ is of no consequence because the sharp momentum dependence of $\Delta R$ cuts off the integral at energies well above this value. Then, expanding the difference of the two distribution functions to first order in the wall velocity, we obtain simple integrals over the reflection asymmetry $\Delta R$. Using the form (45) for $\Delta R$, the resulting 1D-flux is

$$J_L^{1D}/N_c \simeq -\frac{3v\Delta\theta}{4\pi T}A(\xi/2)\,|m|\,w(\xi/2)\frac{e^{2\beta\omega_0}}{(1+e^{\beta\omega_0})^3}\,r_S(w/|m|);$$
$$r_S(w/|m|) = \left(1+2w/|m|+K_1(|m|/2w)e^{|m|/2w}\right). \tag{66}$$

Because the thermal dispersion relations (49) are not Lorentz invariant, it is not straightforward to relate this result to the desired 3-D case. If we assume however that the $k_\|$-dependence of the reflection asymmetry is small, and take a reasonable upper cutoff of $k_{\|,\rm max}=\omega_0$ on the integral over these momenta to ensure that the small-momentum dispersion relations are still valid, we find the result

$$\frac{J_L^{\rm s.m.}}{N_c} \sim -\frac{3v\Delta\theta}{16\pi^2 T}A(\xi/2)\,|m|\,w(\xi/2)\,\omega_0^2\frac{e^{2\beta\omega_0}}{(1+e^{\beta\omega_0})^3}\,r_S(w/|m|) \tag{67}$$

where we used the superscript to distinguish this contribution from that of the large-momentum region (59–62), and the ratio of the three-dimensional phase space to that of one dimension is

$$\int_0^{\omega_0}\frac{d^2k_\|}{(2\pi)^2}=\frac{\omega_0^2}{4\pi}, \tag{68}$$

taking into account the relevant cutoff on the momentum which defines what we mean by the small-momentum region. Using this estimate, we find that ratio of fluxes from the small- to large-momenta regions are approximately

$$\frac{J_L^{\rm s.m.}}{J_L^{\rm l.m.}}=\begin{cases}-0.4, & \tau \text{ lepton;}\\ -6, & b \text{ quark.}\end{cases} \tag{69}$$

It will be shown in section 5.3 that the large-momentum contributions get a suppression of approximately $m/T$ in their contribution to the chiral asymmetry that develops in front of the bubble wall. This being a few percent both for the tau lepton and the $b$ quark, we see that the small-momentum region makes the dominant contribution to baryogenesis. Moreover, as argued in section 4.3, the large-momentum particles approach the wall at a glancing angle and undergo more scatterings with the plasma in the wall, hence their effects will ultimately be even further suppressed.

### 5.1.3 The total flux

To conclude this section we mention that due to the difference of the thermal distribution functions for the left- and right- handed particles, the total flux $J_L + J_R$ is nonzero, as was first pointed out in ref. [34]. We will see that the total flux of leptons is unimportant compared to the left handed flux, but for quarks one must keep track of both. Based on the previous estimate of eq. (59), the contribution to the total flux from the large momentum region is given by

$$J_{\text{tot}}^{\text{l.m.}} \sim N_c \frac{v\Delta\theta Aw|m|^2}{4\pi^2} r_L(w/|m|) \, (E_1^L - E_1^R) \frac{\partial}{\partial E_1}(f(1-f/2)) \tag{70}$$

where $E_1^R$ follows from the definition of $E_1^L$ in (58) by replacing $\omega_L$ by $\omega_R$. For the bottom quark the ratio of the fluxes implied by eqns. (62) and (70) is roughly

$$\frac{J_{\text{tot}}^{\text{l.m.}}(b)}{J_L^{\text{l.m.}}(b)} \simeq -0.1. \tag{71}$$

The total flux coming from the small momentum region vanishes at linear order in the wall velocity, so we must expand the distribution functions of eq. (63) to second order in $vk$, with the result that the total flux is proportional to the previously computed chiral flux (67) according to

$$\frac{J_{\text{tot}}^{\text{s.m.}}}{J_L^{\text{s.m.}}} \simeq 3v\beta\Delta_\omega \frac{e^{\beta\omega_0}-2}{e^{\beta\omega_0}+1}, \tag{72}$$

which gives $\simeq -0.04v$ for the bottom quark. For the taus the total fluxes are clearly ignorably small. For bottom quark however, because total baryon number is conserved by the QCD sphaleron effects and left-handed baryon number is not, as will be seen below, it will turn out that these ratios do not remain small after the reflected quarks interact with the plasma.

## 5.2 Equilibration of Species

Knowing the flux of the two chiralities of a fermion at the bubble wall gives us an initial condition for the problem of how they diffuse into the symmetric phase in front of the wall. During the diffusion process there will be interactions of the fermions with particles in the plasma which change the net fermion densities (the asymmetry between particles and antiparticles), redistributing them amongst other species. For example, interactions with Higgs bosons will convert between the two chiralities. A precise treatment would require the equations for the transport of the two chiralities to be coupled by such interactions. For simplicity we prefer to consider such reactions as either being slow or fast compared to the transport time so that we can deal with uncoupled transport equations. If the reaction times are borderline between these two extremes, we can interpolate between them to get an idea of what the more exact treatment would give. In our model it will turn out to be unnecessary to do so, however.

To decide which interactions are important let us estimate the time scales. The diffusion time scale depends on the bubble wall velocity and the diffusion rate, as will become clear in the next subsection, and is given by $v^2/D$ where $D$ is the diffusion coefficient. This rate is approximately $10^{-3}T$ for quarks at a temperature $T$, and $10^{-4}$ for leptons, assuming a wall velocity of $v = 0.1$ for definiteness. The only interactions with a competitive rate are the strong sphalerons, which are the QCD analog of the usual sphalerons, and the interactions of Higgs bosons with top quarks. The former have a rate around $10^{-2}T$ implied by appropriate scaling of the weak and strong coupling constants [34], and the latter we estimate to be $10^{-3}T$, making certain reasonable assumptions about the Higgs boson masses; the rate of normal sphaleron interactions, by contrast, is $5 \times 10^{-5}T$, consistent with our assumption that it is smaller than the other relevant rates.

Clearly we want to impose the equilibrium of strong sphalerons on our system of fermion asymmetries, conveniently characterized by local chemical potentials for each species. This has important consequences for the quark asymmetries, essentially erasing them up to small corrections (known as mass corrections), although it has no effect on lepton asymmetries. The Higgs-top interactions are marginally in equilibrium on the diffusion time-scale, so we will consider both extremes, when they are approximated as being fast and slow.

Let us introduce chemical potentials, localized at the wall. Since there is no practical difference between the first and second generations, we need only $\mu_{L,R}^u$ and $\mu_{L,R}^d$ to represent up, down, charmed and strange quarks of both chiralities; these will be produced from top and bottom quarks by the strong sphalerons. We also imagine that the tau is the only lepton whose Yukawa coupling is large enough for a significant asymmetry to be produced through reflections at the bubble wall, so that we needn't concern ourselves about the first two generations of leptons. It is also useful to define a quantity $\bar{\mu}$ for each species,

$$\bar{\mu}_i = \mu_i \left(1 - \frac{3m_i^2}{\pi^2 T^2}\right) \equiv \mu_i \left(1 - \delta_i\right), \tag{73}$$

where $m^2$ is the thermal mass of the particle, since we are interested in the symmetric phase. $\bar{\mu}$ is directly proportional to the density of particles [34]-[36], whereas $\mu$ has this property only when the mass corrections are neglected.

We can also define quantities proportional to the densities of various flavor and chiral combinations of baryon and lepton number, as well as weak hypercharge,

$$\begin{aligned}
B^{1+2}_{L(R)} &= 2(\bar{\mu}^u_{L(R)} + \bar{\mu}^d_{L(R)}) \\
B^3_{L(R)} &= (\bar{\mu}^t_{L(R)} + \bar{\mu}^b_{L(R)}) \\
B^{ij}_R &= \bar{\mu}^i_R - \bar{\mu}^j_R \\
L^3_L &= \bar{\mu}^\tau_L + \bar{\mu}^{\nu_\tau}_L \\
L^3_R &= \bar{\mu}^\tau_R \\
Y &= 2(\bar{\mu}^u_L + \bar{\mu}^d_L) + (\bar{\mu}^t_L + \bar{\mu}^b_L) - (\bar{\mu}^\tau_L + \bar{\mu}^{\nu_\tau}_L) \\
&+ 2(4\bar{\mu}^u_R - 2\bar{\mu}^d_R) + (4\bar{\mu}^t_R - 2\bar{\mu}^b_R) \\
&- 2\bar{\mu}^\tau_R + 2n(\bar{\mu}^{H_0} + \bar{\mu}^{H_+}),
\end{aligned} \tag{74}$$

assuming for the moment that there are $n$ Higgs doublets which are in equilibrium with each other. To find the new equilibrium conditions of the chemical species we must impose constraints on the $\mu$'s for each reaction considered to be fast. For the strong sphalerons the condition is

$$2(\mu^u_L + \mu^d_L) + (\mu^t_L + \mu^b_L) = 2(\mu^u_R + \mu^d_R) + (\mu^t_R + \mu^b_R) \tag{75}$$

since they change the chirality of each flavor of quark by two units. The Higgs constraint from interactions with top quarks is

$$\mu^{H_0} = \mu^t_R - \mu^t_L \qquad \text{or} \qquad \mu^{H_+} = \mu^t_R - \mu^b_L. \tag{76}$$

The distinction between the isospin components of weak doublets will turn out to be irrelevant for our results, so that we need not worry about the rate of the weak interactions. The reason is that the weak interaction constraints only serve to determine the chemical potential of the $W^\pm$ bosons, but have no effect on the baryon and lepton asymmetries.

The equilibrium conditions must be solved subject to the constraints that certain quantities are conserved, namely

$$B^{1+2} = B^{1+2}_L + B^{1+2}_R; \qquad B^3 = B^3_L + B^3_R;$$
$$B^{ud}_R = 0; \qquad B^{bu}_R; \qquad L^3_L; \qquad L^3_R; \qquad Y \qquad (77)$$

The last of these is hypercharge; we will deal with the Debye screening of hypercharge in the next subsection. Here let us only note that the list of conserved quantities is augmented with one more, $B^{tu}_R$ say, if the Higgs equilibrium condition (76) is removed.

Our goal now is to solve for the linear combinations of chemical potentials that correspond to total left-handed baryon and lepton number, because it is these that drive the weak sphalerons to make baryons. It is easy to see that if we ignored the mass corrections that distinguish $\mu$'s from $\bar\mu$'s, the left-handed baryon number vanishes due to the strong sphalerons, for then we would have $B^{1+2}_L + B^3_L = B^{1+2}_R + B^3_R$, which coupled with the initial condition that $B^{1+2} + B^3 = 0$ from the reflections would give zero for both chiralities of total baryon number. Actually there is another correction to this statement since, as we mentioned, the net flux of baryon number at the wall is not quite zero due to similar thermal mass corrections. Both effects save the quark reflection asymmetry at the wall from making a vanishing contribution to the final baryon asymmetry. The situation for leptons is considerably simpler: to the order of our approximations, nothing happens to them once they are produced at the wall, aside from the diffusion process which is yet to be considered.

It is a straightforward algebraic task to solve the system of equations for the final chemical potentials in terms of the initial ones. The initial ones are proportional to the fluxes that we computed in the last section; the exact proportionality between fluxes and densities will be discussed shortly. For now we will simply express the final values of left-handed baryon and lepton number, after equilibration has taken place, in terms of conserved quantities, which can be replaced by their initial values. To first order in the thermal mass corrections, the equilibrated values of left-handed baryon number are found to be

$$\begin{aligned}
B_L^{\text{eq}} &= \frac{1}{2}\left(B^{1+2} + B^3 + (\delta_{b_R} - \delta_{t_R})B_R^{bu}\right) \\
&\quad - \left(\delta_{u_L} + \frac{1}{2}(\delta_{u_R} + \delta_{d_R}) + \frac{1}{4}(\delta_{b_R} - \delta_{t_R}) - \delta_{t_R} - \delta_{t_L}\right) \\
&\quad \times \begin{cases} (Y + L_R^3 + (6+4n)B_R^{bu})/(9+14n) & \text{Higgs-top equilibrium;} \\ (B_R^{bu} + B_R^{tu})/3 & \text{no Higgs-top equilibrium.} \end{cases}
\end{aligned} \quad (78)$$

Notice that this equation would vanish if we ignored thermal masses because the total baryon number in each generation is nonzero only due to the thermal masses, as our equation for the total flux of fermion number in the preceding section showed. Now since the right-hand sides are expressed in terms of conserved quantities, we can evaluate them at the initial time, when the asymmetries were produced at the wall, before any equilibration takes place. All the lower generation asymmetries are essentially zero, and hypercharge $Y$ can be expressed in terms of the $B$ and $L$ asymmetries. Furthermore since total $B$ or $L$ is a first-order thermal effect, it can be ignored whenever multiplied by thermal masses as this would be second order. Evaluating the thermal masses,

$$\begin{aligned}
\delta_{u_L} + \frac{1}{2}(\delta_{u_R} + \delta_{d_R}) + \frac{1}{4}(\delta_{b_R} - \delta_{t_R}) - \delta_{t_R} - \delta_{t_L} &= -\frac{3}{\pi^2}\left(\frac{7y^2}{32} + \frac{g'^2}{32}\right); \\
\delta_{b_R} - \delta_{t_R} &= -\frac{3}{\pi^2}\left(\frac{y^2}{8} + \frac{3g'^2}{72}\right),
\end{aligned} \quad (79)$$

and taking $n = 2$ Higgs doublets, the equilibrated value of total left-handed baryon number becomes

$$B_L^{\text{eq}} = 0.5B^3 + \begin{cases} 1.4 \times 10^{-4}(1 - 0.55h_t^2)B_L^b - 0.01B_L^t, & \text{Higgs-top equilibrium;} \\ 1.4 \times 10^{-4}(1 - 2.3h_t^2)B_L^b - 0.04B_L^t, & \text{no Higgs-top equilibrium.} \end{cases} \quad (80)$$

Although we kept the top quark contribution for completeness here, it is practically zero. Furthermore the difference between considering the Higgs-top interactions to be in or out of equilibrium is obviously small for the bottom quark. It should be noted how the left-handed asymmetry is diluted by the equilibrating processes, leaving the initially much smaller total current, conserved by the strong sphalerons, as the dominant source of the injected baryonic asymmetry. Using the previous results (69, 71, 72) for the ratios of injected fluxes, we get

$$\begin{aligned}
B_L^{\text{eq}} &\cong 0.5B^3 = 0.5B_L^b\left(\frac{J_{\text{tot}}^{\text{s.m.}}}{J_L^{\text{s.m.}}}\frac{J_L^{\text{s.m.}}}{J_L} + \frac{J_{\text{tot}}^{\text{l.m.}}}{J_L^{\text{l.m.}}}\frac{J_L^{\text{l.m.}}}{J_L}\right) \\
&= (0.01 - 0.03v)B_L^b,
\end{aligned} \quad (81)$$

where $J_L$ is the sum of the large- and small-momentum contributions. Let us now compare this to the injected $\tau$ lepton current; using $(J_L^{s.m.}(b)+J_L^{l.m.}(b))/(J_L^{s.m.}(\tau)+J_L^{l.m.}(\tau)) = -400$, the ratio is

$$\frac{B_L^{\text{eq}}}{L_L^{\text{eq}}} \simeq -1 + 3v. \tag{82}$$

Therefore we see that the effect of the sphalerons is to reduced the initial preponderance of $b$ quarks over $\tau$ leptons in the injected flux so that they are roughly equal in strength after equilibration. However in the next section we shall see that there is an additional large suppression of the quarks coming from the much larger diffusivity of the leptons, so that in the end the contribution from the quark reflections will be completely overwhelmed by that coming from the $\tau$ lepton reflections. Hence the major conclusion to be drawn from (81) is that the quark reflection is unimportant for the present mechanism of baryogenesis.

Of course for the left-handed lepton flux, which will also bias the sphaleron interactions, we have the trivial relation that $L_L = L_L^\tau$ since under our assumptions the Higgs interactions of the tau lepton are too slow to change its asymmetry. However it has been noted that the Yukawa couplings of the fermions may be larger than we have assumed, since it is possible that the VEV's of the two Higgs fields evolve differently than in our simple model. In this case the tau lepton might have been in equilibrium with the Higgs field on time scales comparable with the diffusion time. The equilibrium conditions would then suppress the final value of left-handed lepton number. By repeating the previous computations with the new equilibrium conditions for Higgs-tau and Higgs-bottom interactions, we find that

$$L_L^3 = L_L^\tau \begin{cases} 1 & \text{no Higgs-tau equilibrium;} \\ (6n+3/2)^{-1} & \text{Higgs-tau equilibrium,} \end{cases} \tag{83}$$

the latter of which cases we include for completeness.

These results can now be used to correct the initial fluxes obtained in the previous section, since we are using the separation of time scales to assume that the initial fluxes at the wall are quickly altered by the establishment of chemical equilibrium before much diffusion into the plasma takes place. We have found that the lepton flux is unaltered (unless the VEV's of the two Higgs fields evolve in a complicated way between the phase transition and now), but that for left-handed quarks is reduced by roughly a factor of 20 from its pre-equilibrium value.

## 5.3 Debye Screening and Diffusion

So far we have computed the initial fluxes from the wall and determined how they are changed by the chemical equilibrium of fast interactions (primarily strong sphalerons) in the plasma. The next step is to propagate the fluxes into the symmetric phase, so see how efficiently they are able to bias the baryon violating interactions of the weak sphalerons.

In the diffusion equation approach it is assumed that the density $n$ of chirality in front of the wall, due to asymmetric reflection of quarks or leptons, is described by the continuity equation and Fick's plus Ohm's law,

$$\partial_t n_i + \partial_z J_i = 0; \quad J_i = -D_i \partial_z n_i + \sigma_i E. \tag{84}$$

Here $D_i$ is the diffusion coefficient for the $i$th particle species, and $\sigma_i$ is its conductivity under the influence of a weak hypercharge electric field which is induced through the particles themselves via Gauss's law, $\partial_z E = \sum y_i n_i$, where $y_i$ is the hypercharge of the $i$th species. It is this coupling between the densities and the gauged charge which gives rise to Debye screening of a certain linear combination of the densities. Eq. (84) and Gauss's law result in coupled equations for the $n_i$,

$$v n_i' + D_i n_i'' - \sigma_i \sum_j y_j n_j = 0, \tag{85}$$

using the fact that for steady state solutions in the rest frame of the bubble wall, $n_i$ has the form $n_i(z - vt)$, so that we can replace the time derivative in (84) by $-v\partial_z$. The general solution of (85) is given in ref. [37]. There it is shown that, for the purpose of computing the baryon asymmetry, one can account for the effect of screening by applying a correction factor $F_i \sim O(1)$ to the solution one would have got by ignoring the screening term in (85),

$$n_i(z) = F_i n_0 e^{-vz/D_i}, \tag{86}$$

where $n_0$ is the density at the wall ignoring screening, to be determined below. As will become apparent, the important quantity for baryon production is the integrated density in front of the wall,

$$\int_0^\infty dz\, n_i(z) = F_i n_0 D_i v^{-1}. \tag{87}$$

This means that the contribution from quark reflections to the baryon reflection will be doubly suppressed compared to that of leptons, as has been emphasized in [8]; once by the strong sphaleron suppression of the flux itself, and again because the diffusion coefficient for quarks is much smaller than that for left-handed leptons: $6/T^2$ versus $110/T$ [8]. In ref. [37] it was shown that $F_l = 1.75$ for the left-handed leptons in the case that the quarks are completely neglected. Moreover we checked that including the quark fluxes, as computed from the results of section 5.2, would produce only a minor change which could be accounted for by taking $F_l \simeq 1.85$ instead. Since the quark contribution is further suppressed by thermal damping (see section 4.3), it is clearly neglible in comparision to the contribution of tau lepton reflection.

We must now determine the chiral density at the wall, $n_0$, in terms of the chiral flux injected at the wall, $J_L$, which was computed in section 5.1. To do so, we will imagine that the wall deposits an infinitesimal amount of chiral density at each point in space $x_i$ when it passes by at time $t_i = x_i/v$, which at first is localized exactly and then spreads out in accordance with the diffusion equation. Integrating all these contributions gives the total chiral density due to the flux of particles reflected from the wall:

$$N(x - vt) = c \int_{-\infty}^{vt} dx_i \, \frac{e^{-(x-x_i)^2/4D(t-t_i)}}{\sqrt{t-t_i}} = c \int_0^\infty dz \, \frac{e^{-v(x-vt+z)^2/4Dz}}{\sqrt{z}}. \tag{88}$$

Notice that (88) is a solution to the diffusion equation (85), ignoring the screening term, but with a delta function source localized at $x = vt$, the position of the wall. The second form comes from the changes of variables $x_i = z + vt$, and the constant of proportionality $c$ can be determined by conservation of particle number. To do so, we note that for $x - vt \to \infty$, the expression (88) approaches an asymptotic value $N_0 = 2c\sqrt{\pi D/v}$. Since the wall is moving at velocity $v$, the rate at which the chiral charge is being created per unit area is $N_0 v$, and this must be equal to the flux $J_L$ injected from the wall.

However this is not yet the correct identification of $n_0$ in eq. (86) because we must remember that there is an equal and opposite chiral flux being injected by the wall in the opposite direction. However the diffusion equation does not "know" it is being injected in the opposite direction; it only knows that it has the opposite sign. If we were to add the two contributions naively, they would exactly cancel each other. What must happen, in fact, is that the two fluxes penetrate a distance $\Delta_P$ into the plasma before they become thermalized and the diffusion equation becomes a valid description [8]. Therefore the correct expression for the chiral density is

$$n(z) = N(z + \Delta_P) - N(z - \Delta_P), \tag{89}$$

and if $\Delta_P \ll D/v$, its integrated value is approximately given by

$$\int_0^\infty dz\, n(z) = 2J_L \Delta_P/v, \tag{90}$$

which must still be corrected with the factor $F_i$ to account for hypercharge screening as in eq. (87). This is precisely the result one would get by putting the source term $2J_L\Delta_P \delta'(z) \cong J_L(\delta(z+\Delta_P) - \delta(z-\Delta_P))$ into the diffusion equation, as was done in ref. [8]. We believe that the present derivation clarifies their procedure.

---

[2]The value $6/T$ agrees with an independent calculation done in ref. [33]. It can be shown the the momentum-space diffusion coefficient $\tilde{D}$ computed there is related to the normal one by $D = T^2/\tilde{D}$.

For the thermalization distance $\Delta_P$, we take the estimate made in ref. [8] of the distance over which scatterings in the plasma will randomize the velocities of the particles in the injected flux (this might underestimate the injection distance for the low momentum states, but we nevertheless use it as a conservative estimate):

$$\Delta_P = 3Dv_i, \tag{91}$$

where $v_i$ is the average velocity of the particles in $J_L$. Following the logic of section 5.1, this velocity is the ratio of the flux, eq. (59), to the same expression except without the factor of $k_z/E_1$ in the integrand. For the large-momentum region of phase space discussed in the section 5.1 the result is

$$v_i = \left[\frac{r_L(a)}{r_S(2a)}\right]\left[\frac{m}{E_1^L + T(1+e^{\beta E_1^L})\ln(1+e^{-\beta E_1^L})}\right], \tag{92}$$

where $a = w/m$ and the other symbols are defined in eq. (60) and (66). For the tau lepton, this gives $v_i = 2.4m/T$, while for the $b$ quark it is $v_i = 1.3m/T$. For particles injected at small momentum, the direction is nearly perpendicular to the wall and the dispersion relation (49) implies

$$v_i = 1/3. \tag{93}$$

# 6 The Baryon Asymmetry

With the preceding results it is a simple matter to compute the baryon asymmetry, because the rate of baryon violation due to sphalerons in terms of the total left-handed quark and lepton densities is

$$\dot{n}_B = -9(\Gamma/T^3)(n_{q_L} + n_{l_L}). \tag{94}$$

This can easily be derived from the Boltzmann equation in the following way. Define forward (backward) sphaleron interactions as those which change baryon number by $+3$ $(-3)$ units. Half of the time the sphaleron interaction will involve one member of a given fermion doublet and half the time the other; if we ignore the distinction between the two at first and then average over different doublet members at the end we will get the right answer. Since baryon number is violated by 3 units, the Boltzmann equation is

$$\dot{n}_B = 3\sum \int d\Pi \left( f_1 \cdots f_n(1-f_{\bar{1}})\cdots(1-f_{\bar{n}}) - f_{\bar{1}}\cdots f_{\bar{n}}(1-f_1)\cdots(1-f_n) \right) \tag{95}$$

where the sum is over all possible channels, the integration measure includes the squared matrix element and the delta function for 4-momentum conservation, and the $f$'s are Fermi-Dirac distribution functions for $n$ ($\bar{n}$) initial (final) states. The Pauli blocking factors can be written as

$$1 - f_i = e^{\beta(E_i-\mu_i)} f_i \cong e^{\beta E_i}(1 - \beta\mu_i) \tag{96}$$

so that (95) becomes

$$\dot{n}_B = 3\beta \sum (\mu_{\bar{1}} + \cdots + \mu_{\bar{n}} - \mu_1 - \cdots - \mu_n) \int d\Pi f_1 \cdots f_n f_{\bar{1}} \cdots f_{\bar{n}} e^{\beta(E_1+\cdots+E_n)}, \tag{97}$$

using energy conservation to equate $E_1 + \cdots + E_n$ to $E_{\bar{1}} + \cdots + E_{\bar{n}}$. The combination of chemical potentials appearing here is always the same one no matter what channel, once we average over the members of the doublets: it is half the sum of potentials for 18 left-handed quarks and 6 left-handed leptons associated with the sphaleron. The sum over channels of the integral is by definition the rate of sphaleron interactions per unit time and volume. Using the fact that density is related to the chemical potential by $n = \mu T^2/6$ for a single lepton flavor and $n = \mu T^2/2$ for a single quark flavor (because of the three colors), we arrive at (94) after summing over generations and averaging over members of doublets.

We have shown that $n_q \ll n_l$ in front of the wall, so that only the lepton contribution need be considered. The interaction rate per unit volume $\Gamma$ is $\kappa_{\text{sph}}(\alpha_W T)^4$ in the symmetric phase. To find the density of baryons produced by sphalerons at a given position $z$, one integrates (94) from $t = -\infty$ until the time when the bubble wall passes the position $z$, when the sphaleron interactions effectively turn off. The integral over time can be rewritten as an integral over distance in front of the wall by a change of variables:

$$n_B = -\frac{9\Gamma}{T^3 v}\int_0^\infty dz\, n_{l_L}(z). \tag{98}$$

A convenient measure of the baryon asymmetry is the ratio of $n_B$ to the entropy density of the universe, $s = 2\pi^2 g_* T^3/45$, with $g_* = 110.75$ degrees of freedom at the electroweak phase transition. Then, assembling our previous results, we have that

$$\frac{n_B}{s} = \frac{1215\,\alpha_W^4\,\kappa_{\text{sph}}}{\pi^2 g_* T_c^2 v^2} D_l F_l (v_i J_L + \tfrac{1}{3} J_L^{\text{s.m.}}), \tag{99}$$

recalling that the expressions for the fluxes were given in eqs. (59) and (67). Putting in the numbers pertinent for the tau lepton contribution $D_l = 110/T$, $F_l = 1.85$, $A(0.06) = e^{-3.2}$, $w(0.06) = 2.6m$, $r_S(0.12) = 12.4$, $\omega_0/T = 0.17$ and $m(T_c)/T_c = 0.01$), and using the recent result $\kappa_{\text{sph}} = 1.1$ corresponding to the classical sphaleron transition rate [39], we obtain

$$\frac{n_B}{s} = 1 \times 10^{-12}\,\frac{\Delta\theta}{v} \tag{100}$$

Given the range allowed by primordial nucleosynthesis: $n_B/s = 1.4 - 3.8 \times 10^{-11}$ [42], eq. (100) translates to a constraint for the parameters $\Delta\theta$ and $v$:

$$15 \lesssim \frac{\Delta\theta}{v} \lesssim 40. \tag{101}$$

Although recent estimates [40, 41] predict rather small terminal velocities corresponding to deflagrating bubbles, $v \sim 0.3$, this is still too large to satisfy eq. (100), even if $\Delta\theta = 1$. However it is possible to imagine means by which the final asymmetry could be boosted to the desired level, as we now discuss.

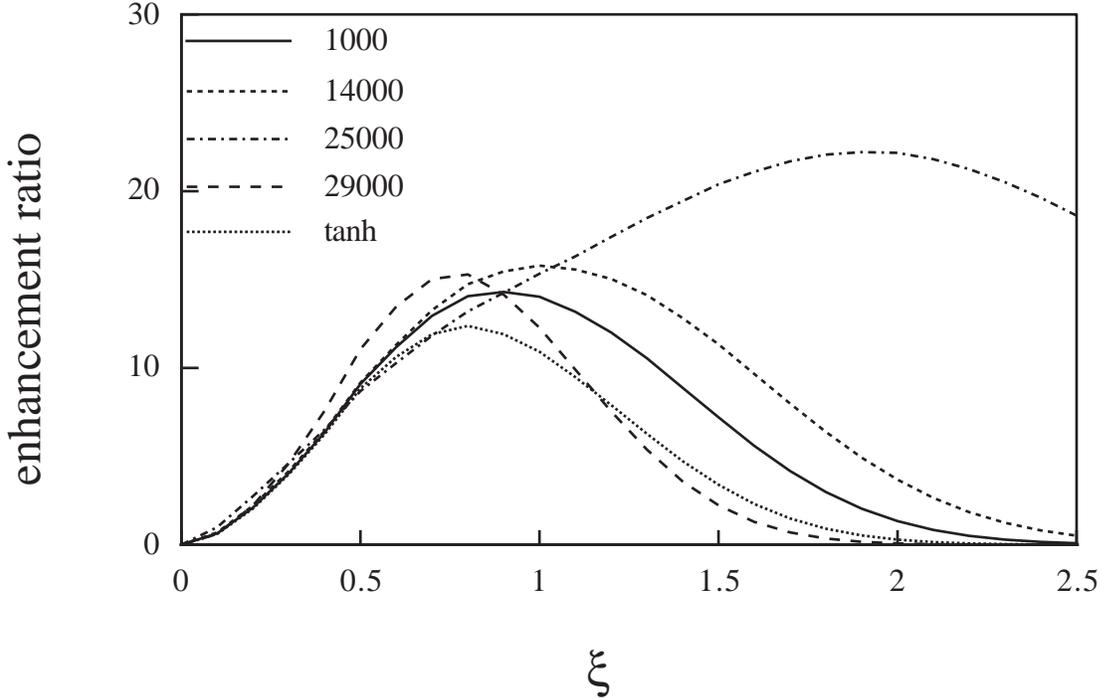

**Figure 6.** The enhancement of the generated baryon asymmetry as a function of the effective fermion mass parameter $\xi = m\Delta_{\text{wall}}$ (see eq. (48)). The curve is normalized to unity at the value $\xi = 0.12$ which we take to correspond to the tau lepton. As in figure 3, the curves are labeled by the corresponding value of $-\kappa$ in GeV$^2$.

A very promising possibility of enhancement is provided by a dynamical slowing down of the bubble walls due to the heating of the plasma in the unbroken phase by the shock waves of the neighbouring bubbles [40]. This deceleration always occurs for deflagration bubbles, and while it has a strong quantitative dependence on the dynamical details of the transition, it is qualitatively easy to understand: the heating of the unbroken phase reduces the difference of the free energies between the interior and exterior regions of a bubble, which is the driving force of the expansion, so the walls slow down when eventually hit by the shock waves of the neighbouring bubbles. Then, given that the wall velocity goes down by a large factor when, say, half of the universe is still in the unbroken phase, and knowing that the baryon production rate goes like $1/v$, it becomes evident that essentially all baryons might have been produced in this later, decelerated phase of the transition. Ref. [40] suggests the possibility of a very large deceleration, $v \to v/100$, which would easily make the present mechanism a viable candidate for baryogenesis. We stress however the sensitive dependence of this effect on the dynamics of the transition and that we have not attempted to calculate its size in the present model.

Another way of increasing the above result was suggested in ref. [8], namely to increase the ratio $m/T_c$ during the phase transition beyond its value in our model, by invoking possible complications in the evolution of the two Higgs fields: if the one that couples to the tau lepton has a larger VEV relative to the other Higgs field during the phase transition than at zero temperature, then the power-law dependence on $m/T_c$ would boost the production of baryons. This corresponds to taking a larger value of the dimensionless parameter $\xi = m\Delta_{\text{wall}}$ than that ($\xi = 0.12$) which our model gave. We have explored the dependence of the baryon asymmetry on $\xi$ and summarized the results in figure 6. Although the enhancement depends on the details of the bubble wall profile, which in turn depends on the Higgs potential parameter $\kappa$, for most values of $\kappa$ the optimal fermion mass occurs in the vicinity of $\xi = 1$, corresponding to one inverse bubble wall width. Thus the mass of the tau lepton is not very far from being the ideal size given the width of the wall in our model.

Figure 6 shows that one can get somewhat larger values if $\kappa$ is tuned to particular values, as the case of $\kappa = -25000$ GeV$^2$ illustrates. It is possible to understand this enhancement qualitatively [22]. In the classical limit of the fermion scattering off the wall, the gradient of the $\theta$-field acts like an effective potential to be added on top of the usual wall potential, with opposite sign for particles and antiparticles. Therefore if $\theta$ is mainly changing well inside the broken phase, some particles whose momentum would otherwise get them over the barrier effectively see a 'bump' that causes them to be reflected. At the same time antiparticles with the same momentum see no such bump and are transmitted. Although quantum mechanics will reduce this effect because of tunneling, one nevertheless expects to see an enhancement in the difference $\Delta R$ of the reflection probabilities. Comparison with figures 1 and 3 shows that, in contrast to the other cases, where $\partial\theta/\partial z$ is concentrated toward the front of the wall, in the case of $\kappa = -25000$ GeV$^2$, $\theta(x)$ is indeed changing primarily within the broken phase.

# 7  Results and Conclusions

We have attempted to make a quantitatively accurate estimate of the baryon asymmetry in the charge transport mechanism of electroweak baryogenesis, using a somewhat realistic two-Higgs doublet model. By assuming reasonable values of parameters, we find that the tau lepton is by far the most important particle species contributing to the baryon asymmetry through its CP-violating reflections from the domain walls that form during the phase transition, and that the resulting baryon asymmetry can be marginally big enough for consistency with primordial nucleosynthesis.

In this section we will remind the reader of our assumptions and try to indicate how our conclusions depend upon them.

1) Concerning the phase transition, we tuned the parameters of the Higgs potential to give $\tan\beta = 1$ for the ratio of the two Higgs field VEV's, because we chose not to deal with a two-stage phase transition, in which one field gets a VEV before the other does. In order for the fermionic loop corrections not to spoil this tuning, that is, to keep also the temperature-dependent Higgs masses equal, it was necessary to couple heavy fermions with equal strength to both Higgs fields. However, if we let *all* fermions couple to each Higgs field with exactly the same strength, then there would be no CP-violation in the fermion mass, because the two fields would give contributions with canceling imaginary parts. In this paper we assumed equal couplings of the top quark to the Higgs fields in the effective action, yet computed the quark reflection as if they only coupled to one of the fields. It should be clear that our conclusion about the smallness of the asymmetry arising from quark reflections does not depend on these details. So in retrospect we see that we can choose to couple quarks symmetrically to the Higgs bosons yet have the leptons coupling only to $\Phi_2$. In such a model the transition truly proceeds simultaneously in both Higgs fields, and our treatment is completely self-consistent. Moreover we believe that our results are representative also of the two stage phase transitions, because as long as one keeps $\rho_c/T_c = 1$ to satisfy the sphaleron washout constraint, the baryon asymmetry should not change much since the ratio to which it is most sensitive, $m(T_c)/T_c$, remains constant.

2) The problem of sphaleron washout also prompted us to assume a small value of the

mass of the lightest neutral Higgs particle, with a definite corresponding value of the effective quartic coupling $\lambda_{\text{eff}} = 0.12$. The ratio $\rho_c/T_c$ decreases with increasing $\lambda_{\text{eff}}$, making the sphaleron interactions not sufficiently suppressed in the broken phase inside the bubbles. But this ratio also depends on the cubic term in the high-temperature Higgs potential. We had to assume a larger value of the cubic coupling than predicted within our model in order to avoid the washout problem. The situation is ameliorated somewhat by recent work [43] which finds a suppression of the sphaleron rate inside the bubbles due to loop effects. Moreover, there is a large number of cubic-like contributions coming from the scalar fields, which we omitted because of technical reasons, that might tend to increase the effective cubic term in this model. Finally one might expect that nonperturbative effects in the symmetric phase play the same role in the two-doublet model as has been recently found in the standard model [20], increasing the amount of supercooling and hence effectively increasing the ratio $\rho_c/T_c$. This phenomenon should also be roughly mimicked by a larger effective cubic term.

3) We treated the CP-violating phase $\theta(x)$ as a perturbation which had no back-reaction on the VEV $\rho(x)$ of the Higgs fields. Since there are no strong constraints on this phase, and none at all if it arises spontaneouly at finite temperature, this assumption was not necessary and served only as a convenience. It is possible that a complete solution of the coupled equations for $\rho(x)$ and $\theta(x)$ would give different results, but if the $\kappa$-dependent changes of the shape of these solutions give any indication (figure 3), we do not expect much sensitivity except for fermion masses significantly larger than the inverse wall thickness.

If the tau lepton was heavier by a factor of 5, the baryon production in the present mechanism would be increased by a factor of $15-80$, making it a viable mechanism of baryogenesis. One obviously cannot change this fact of nature, but we can imagine ways of making $m_\tau$ effectively heavier during the phase transition. One possibility would be to actually find a model that displays the behavior suggested by ref. [8] in which the VEV's of the two Higgs fields obey $\rho_1(T_c)/\rho_2(T_c) > \rho_1(0)/\rho_2(0)$. One should in this case also take into account other effects of a two-stage phase transition on baryogenesis. A second possibility we discussed is that at the final stage of the phase transition the bubble walls are slowed down due to the heating of the unbroken phase by the shock waves of the neighbouring bubbles, which could lead to a large enhancement of the baryon production. A third possibility would be to demonstrate in a more convincing fashion that the cubic term in the effective potential could be increased even more than we assumed above, so that $m(T_c)/T_c$ would be increased. This would also be welcome from the point of view of insuring that sphaleron interactions in the broken phase are too slow to destroy the baryon asymmetry that nature may have so intricately produced at the electroweak phase transition.

## Acknowledgements

This research is supported by DOE grants DE-AC02-83ER40105 and DE-FG02-87ER40328, N.S.E.R.C. of Canada, and les Fonds F.C.A.R. du Québec. K.K. would like to thank M. Shaposhnikov for enlightening discussions.

## A  Analytic solution to the linearized $\theta$-equation

In this appendix we derive an analytical solution to the linearized equation of motion for $\theta(x) = \theta(g(x))$, eq. (28). Recall that $g(x)$ describes the modulus of the Higgs field at the bubble wall, eq. (21).

$$4g^2(1-g)^2\frac{\mathrm{d}^2\theta}{\mathrm{d}g^2} + 4g(1-g)(3-4g)\frac{\mathrm{d}\theta}{\mathrm{d}g} + (B+Cg^2)\theta = -D \tag{102}$$

The two homogeneous solutions to this equation are given by

$$\begin{aligned}\theta^+(g) &= \frac{1}{g}g^{-\alpha}(1-g)^{+\beta}\,_2F_1(-\alpha+\beta+\frac{1}{2}+\gamma,-\alpha+\beta+\frac{1}{2}-\gamma;1+2\beta;1-g),\\ \theta^-(g) &= \frac{1}{g}g^{+\alpha}(1-g)^{-\beta}\,_2F_1(+\alpha-\beta+\frac{1}{2}+\gamma,+\alpha-\beta+\frac{1}{2}-\gamma;1+2\alpha;g)\end{aligned} \tag{103}$$

where the $_2F_1$ are hypergeometric functions characterized by the parameters

$$\begin{aligned}\alpha &= \sqrt{1-B/4}\\ \beta &= \frac{1}{2}\sqrt{-B-C}\\ \gamma &= \frac{1}{2}\sqrt{9-C}.\end{aligned}$$

The homogeneous solutions $\theta^\pm$ diverge for $x \to \pm\infty$. But our boundary conditions require finiteness in these limits. We therefore have to set the coefficients of the homogeneous solutions (103) identically to zero and are left with the inhomogeneous solution.

The inhomogeneous solution can be constructed by use of the Greens function of equation (102). We find

$$\theta(x) = -\frac{1}{4}DW\left[\theta^+(g(x))\int_0^{g(x)} dg' \frac{g'}{1-g'}\theta^-(g') + \theta^-(g(x))\int_{g(x)}^1 dg' \frac{g'}{1-g'}\theta^+(g')\right] \quad (104)$$

where the Wronskian $W$ is given by

$$W = \frac{\Gamma(\alpha+\beta+\frac{1}{2}+\gamma)\Gamma(\alpha+\beta+\frac{1}{2}-\gamma)}{\Gamma(1+2\alpha)\Gamma(1+2\beta)}. \quad (105)$$

Although we cannot further simplify (104), we can find an approximation to (104) in the thick wall limit. In this adiabatic case one would expect that the kinetic term is irrelevant throughout. We then simply ignore the kinetic term in (102) and find the solution for $\delta V/\delta\theta(x) = 0$.

$$\theta_{\text{adiab}}(x) = \frac{-D}{B+Cg^2}. \quad (106)$$

This solution can be verified using (104).

## B  Finite temperature Dirac equation

In this appendix we outline the derivation of the Dirac equation for the scattering of a fermionic excitation off a bubble wall including the effects of the thermal background. While most of the equations shown below have been derived elsewhere, we present them here for completeness and in order to be able to discuss their implications for the present physical application.

The basic computational task is to compute the thermal self-energy corrections to the fermion propagator. In the unbroken phase chirality is a good quantum number, so that the self-energy separates in the chiral representation. Going directly to the rest frame of the plasma, one may show that [25]

$$\Sigma_X = \left\{(a_X(k,\omega)\omega + b_X(k,\omega))\gamma^0 - a_X(k,\omega)\mathbf{k}\cdot\boldsymbol{\gamma}\right\}P_X, \quad (107)$$

where $X = L, R$ refers to the chirality, $P_X$ is the corresponding chiral projector, $\mathbf{k}$ and $\omega$ are the three-momentum and the energy of the particle in the plasma rest frame and the functions $a$ and $b$ have the well-known high temperature ($k, \omega \ll T$) limit [25],

$$\begin{aligned}
a_X(k,\omega) &= \frac{\omega_X^2}{k^2}\left(1 - \frac{\omega}{2k}\ln(\frac{\omega+k}{\omega-k})\right) \\
b_X(k,\omega) &= \frac{\omega_X^2}{k}\left(\frac{\omega}{k} - \frac{\omega^2-k^2}{2k^2}\ln(\frac{\omega+k}{\omega-k})\right).
\end{aligned} \quad (108)$$

In the approximation that the masses of the particles within the loops can be neglected, the above expressions are valid also in the broken phase (and within the wall). Then the effective Dirac equation, at the one loop level becomes

$$\begin{pmatrix} (1-a_L)\omega - b_L + (1-a_L)\sigma\cdot\mathbf{k} & -m \\ -m^* & (1-a_R)\omega - b_R - (1-a_R)\sigma\cdot\mathbf{k} \end{pmatrix}\begin{pmatrix} L \\ R \end{pmatrix} = 0. \quad (109)$$

Because of the nonlinear dependence of the functions $a$ and $b$ on the energy and momentum, this is a highly nonlocal equation, which is a reflection of its inherent multiparticle nature. In particular the nonlinearity in energy makes it impossible to give it an exact interpretation in terms of effective single particle states, except in the small and large momentum limits, where the self-energy can be approximately linearized and the (re)quantization procedure may be completed.

In order to proceed with the reflection computation however, one needs to find at least an approximate interpretation of (109) in terms of single particle states. The remedy is of course well known; one defines the effective quasiparticle states as the collective excitations corresponding to the poles of the 1-loop propagator, or in other words, to the peaks in the phase space density in energy. One should however bear in mind that such an interpretation does not give a complete description of the system and in some cases pushing the picture too far can lead to ambiguities.

The poles of the propagator correspond to the zeros of the determinant of the matrix appearing in (109). In the symmetric phase the resulting dispersion relations take the particularly simple form

$$g_X^{\text{p,h}}(\omega, k) \equiv (1-a_X)(\omega \mp k) - b_X = 0, \quad (110)$$

where the two signs correspond to two different branches of solutions: the one with the minus sign can be viewed as a generalization of the usual particle excitation to finite temperatures.

The one with the plus sign on the other hand represents a new solution that has no counterpart at zero temperature wave function renormalization. Indeed by constructing a 1-particle propagator [27, 26]. The appearance of this new 'hole' excitation apparently leads to unphysical doubling of the number of degrees of freedom (as measured by the volume of phase space), which calls for and is corrected by finite-temperature wave function renormalization. Indeed by constructing a 1-particle propagator from the quasiparticle states and comparing to the full propagator, one finds that the correctly normalized quasiparticle wave functions differ from the vacuum wave functions by the momentum-dependent normalization factor $(Z_X^{\mathrm{p,h}}(k))^{1/2}$, where [25, 27, 44]

$$Z_X^{\mathrm{p,h}}(k)^{-1} = \left(\frac{\mathrm{d}g_X^{\mathrm{p,h}}}{\mathrm{d}\omega}\right)_{\omega=\omega^{\mathrm{p,h}}(k)} = \left[1 + a_X + (1 - a_X)\frac{\omega \mp k}{\omega \pm k}\right]_{\omega=\omega^{\mathrm{p,h}}(k)}. \tag{111}$$

One can readily work out the limiting values of $Z$: in the small momentum limit

$$Z_X^{\mathrm{p,h}}(k) \simeq \frac{1}{2} \pm \frac{k}{3\omega_X} \tag{112}$$

and in the large momentum limit (where also $\omega \simeq k$)

$$\begin{aligned} Z_X^{\mathrm{p}}(k) &\simeq 1 - \frac{\omega_X^2}{k^2}\ln\frac{\omega_X}{k} \\ Z_X^{\mathrm{h}}(k) &\simeq e^{-2k^2/\omega_X^2 - 1}. \end{aligned} \tag{113}$$

Thus, in the small momentum limit both particle and hole excitations have equal weight, which is half of the zero temperature value. Moreover, holes are only present at momenta $k \lesssim \omega_X$, above which their effective number density (given by $Z(k)f(\omega(k))$) falls off exponentially. Therefore one does not need to account for holes in the high-momentum region.

The small momentum limit dispersion relations (49) and the Dirac equation (51) are easily derived from (109) and (110) after finding the small-momentum limits of the functions $a$ and $b$:

$$a_X = \frac{\omega_X^2}{3\omega^2} + \mathcal{O}(k^2); \qquad b_X = 2a + \mathcal{O}(k^2). \tag{114}$$

The factor of $1/2$ coming from $Z_X^{\pm}(0)$ for both particles and holes of both chiralities, which compensates the doubling of the number of excitations due to the appearance of the hole states, was included in our equation (63). This factor was overlooked in the treatments of the quasiparticle scatterings in the standard model in references [26, 29].

In the large momentum limit the hole excitations vanish, since their wave function renormalization factor goes exponentially to zero. Moreover, since both $a$ and $b$ go to zero at large $k$,

$$a_X \simeq \frac{\omega_X^2}{k^2} \ln \frac{k}{\omega_X}; \qquad b_X \simeq \frac{\omega_X^2}{k}, \tag{115}$$

one might expect that the Dirac equation trivially approaches the vacuum equation. The situation is more complicated however, because we are interested in phenomena that depend on small differences between energy and momentum. In fact one can show that the $a$-factor may safely be neglected, but that the remaining equation has other terms that are of the same order as the $b$-term even at high momentum. Nevertheless, one would still expect that the vacuum equation gives a reasonable approximation for the reflection of high-momentum particles, since the $b$-term affects both symmetric and broken phases equally.

Let us finally point out that in the intermediate momentum region $k \sim \omega_X$ the wave function normalization factors do not add up to 1; instead their sum can be as low as about 0.8 [27]. This signals the breakdown of the single particle interpretation, which can lead to inconsistencies. For instance, replacing the limiting normalization factors $\frac{1}{2}$ by corresponding momentum-dependent $Z$'s leads to a small nonvanishing flux in the intermediate momentum region, even when the wall is not moving. However, this flux is much smaller than, and clearly caused by, the inherent error in the total flux due to the abovementioned fact that in this region the effective 1-particle states do not give a faithful representation of the phase space.